\documentclass[twocolumn, twocolappendix]{aastex631}

\usepackage{graphicx}
\usepackage{amsmath}% http://ctan.org/pkg/amsmath
\usepackage{comment}
\usepackage{newtxtext,newtxmath}

\usepackage{float}
\usepackage{pifont}% http://ctan.org/pkg/pifont
\usepackage{enumerate}

\shorttitle{The edges of intracluster light in simulated clusters}
\shortauthors{Dacunha et al.}
\begin{document}

\title{Memoirs of mass accretion: probing the edges of intracluster light in simulated galaxy clusters}
\author[0000-0002-4746-2128]{Tara Dacunha}
\affiliation{Kavli Institute for Particle Astrophysics \& Cosmology, P. O. Box 2450, Stanford University, Stanford, CA 94305, USA}
\affiliation{Department of Physics, Stanford University, 382 Via Pueblo Mall, Stanford, CA 94305, USA}
\affiliation{SLAC National Accelerator Laboratory, Menlo Park, CA 94025, USA}
\author[0000-0001-9863-5394]{Phil Mansfield}
\affiliation{Kavli Institute for Particle Astrophysics \& Cosmology, P. O. Box 2450, Stanford University, Stanford, CA 94305, USA}
\affiliation{SLAC National Accelerator Laboratory, Menlo Park, CA 94025, USA}
\author[0000-0003-2229-011X]{Risa H. Wechsler}
\affiliation{Kavli Institute for Particle Astrophysics \& Cosmology, P. O. Box 2450, Stanford University, Stanford, CA 94305, USA}
\affiliation{Department of Physics, Stanford University, 382 Via Pueblo Mall, Stanford, CA 94305, USA}
\affiliation{SLAC National Accelerator Laboratory, Menlo Park, CA 94025, USA}

\email{tdacunha@stanford.edu}

\begin{abstract}
The diffuse starlight extending throughout massive galaxy clusters, known as intracluster light (ICL), has the potential to be read as a memoir of mass accretion: informative, individual, and yet imperfect. Here, we combine dark matter-only zoom-in simulations from the Symphony suite with the \textsc{Nimbus} ``star-tagging'' model of the stellar halo to assess how much information about the mass assembly of an individual galaxy cluster can be gleaned from idealized measurements of ICL outskirts. We show that the edges of a cluster's stellar profile---the primary ($R_{\textrm{sp}\star,1}$) and secondary ($R_{\textrm{sp}\star,2}$) stellar ``splashback'' radii---are sensitive to both continuous mass accretion histories and discrete merger events, making them potentially powerful probes of a cluster's past. We find that $R_{\textrm{sp}\star,1}$ strongly correlates with the cluster's mass $\sim$1 dynamical time ago, while $R_{\textrm{sp}\star,2}$ traces more recent mass accretion history to a slightly lesser degree. In combination, these features can further distinguish between clusters that have and have not undergone a major merger within the past dynamical time. We use both to predict realistic cluster mass accretion histories with the \textsc{MultiCAM} framework. These outer ICL features are significantly more sensitive to mass accretion and merger histories than the stellar mass gap and halo concentration, and perform comparably to the commonly used X-ray-based tracer of relaxedness, $x_{\rm off}$. While our analysis is idealized, the relevant ICL features are potentially detectable in next-generation deep imaging of nearby clusters. This work highlights the promise of ICL measurements and lays the groundwork for more detailed forecasts of their power.
\end{abstract}

\keywords{galaxies: clusters: general --- galaxies: formation --- galaxies: halos --- dark matter --- cosmology: theory}

%%%%%%%%%%%%%%%%%%%%%%%%%%%%%%%%%%%%%%%%%%%%%%%%%%

%%%%%%%%%%%%%%%%% BODY OF PAPER %%%%%%%%%%%%%%%%%%

\section{Introduction}
\label{sec:intro}

Galaxy clusters are enveloped in a vast and diffuse cloud of starlight known as the intracluster light (ICL), composed of stars that are gravitationally bound not to individual member galaxies, but to the cluster potential itself.
Built up through catastrophic major mergers as well as the tidal stripping and disruption of smaller satellite galaxies (see \citealt{Contini_2021} and \citealt{Montes_2022} for recent reviews), the ICL carries a cumulative record of a cluster’s past members and accretion history. 
As the ICL's stars are ex-situ, collisionless, and follow the gravitational potential, the ICL may further serve as a luminous counterpart to the dark matter. Thus, the ICL offers a unique opportunity to trace the vast spatial extent of dark matter halos (e.g. \citealt{Montes_Trujillo_2019, Asensio_Vecchia_Bahé_Barnes_Kay_2020, Yoo2022, Yoo2024}) as well as key signatures of dark matter structure formation using light.

In dark matter halos, a particularly informative dynamical feature is the splashback radius---the location of the collective first apocenters of recently accreted material. The location of these apocenters forms a quasi-spherical shell around the halo (e.g., \citealt{Adhikari2014, Mansfield_2017}) and leads to a sharp drop in local density. The splashback radius corresponds to a transition between purely infalling and virialized orbits, providing a physically motivated boundary of the halo. Notably, this radius has been shown to correlate strongly with mass accretion rate and discrete merger events due to the contraction of particle orbits in a rapidly deepening potential (e.g., \citealt{Adhikari2014}, \citealt{DK14}; \citealt{More2015}, \citealt{Zhang_Zhuravleva_Kravtsov_Churazov_2021}, \citealt{Shin_Diemer_2022}). This dynamical feature need not be confined to dark matter; the collisionless stars that form the ICL analogously accrete onto and orbit the same gravitational potential well.

A luminous reflection of the splashback feature could bring to light the mass accretion history (MAH) and the dynamical state of individual clusters, allowing for better understanding of scatter in scaling relations and cluster properties, the identification of merging systems for analysis or exclusion in equilibrium studies, and deeper understanding of how these massive systems form and evolve.
\citet{Deason_StellarSplashback} found that simulated clusters exhibit two splashback-like features in the stellar halo, corresponding to the first and second orbital apocenters of stars and related to the dark matter splashback feature and recent mass accretion rate. Observationally, \citet{Gonzalez_George_Connor_Deason_Donahue_Montes_Zabludoff_Zaritsky_2021} subsequently presented tentative evidence for a steep drop in the ICL of MACS J1149.5+2223, potentially associated with the stellar splashback radius. These studies raise the intriguing possibility that the outer ICL can be used to trace not only a cluster's extent, but its rich assembly history. 

In recent years, several other features of the ICL and stellar halos have emerged as probes of the recent merger histories of clusters and massive galaxies. These features have included stellar mass density profiles \citep[e.g][]{Pillepich_Vogelsberger_Deason_Rodriguez-Gomez_Genel_Nelson_Torrey_Sales_Marinacci_Springel_etal._2014,Rey_Starkenburg_2021, Genina_Deason_Frenk_2022, Yoo2024}, two-dimensional spatial distributions of the ICL compared to the dark matter \citep[e.g.][]{Yoo2022, Yoo2024}, and the fraction of stellar mass residing in the ICL \citep[e.g.][]{Conroy2007, Rudick2011, Cui2014, Contini2023, ContrerasSantos2024, GoldenMarx2024_maggap, Kimmig2025}. 

Despite the emergence of the ICL as a potential tracer of merger histories as well as dark matter halos and the tantalizing hints of dynamical outer ICL features, the predictive power of stellar splashback features has yet to be fully explored. Do these features encode measurable signatures of a cluster’s past accretion and merger activity? How do they compare to more traditional proxies of formation history, such as halo concentration or the satellite stellar mass gap? And can they be used to reconstruct continuous mass accretion histories of individual clusters?

In this paper, we probe how much information in an individual cluster’s mass accretion and merger history can be extracted from the stellar splashback features of its ICL. In other words, we read the faint starlight of galaxy clusters as a memoir of mass accretion: informative, individual, but likely imperfect. Using high-resolution Symphony simulations and the \textsc{Nimbus} star-tagging model, we make profile- and orbit-based estimates of the primary and secondary stellar splashback radii and analyze their correlation with continuous and discrete mass accretion histories. We assess their predictive power using the \textsc{MultiCAM} framework, and compare their performance to established structural and X-ray-based tracers. Finally, we consider their observational prospects, and the promise of ICL splashback measurements in next-generation surveys.

\begin{figure}[t]
\includegraphics[width=\columnwidth]{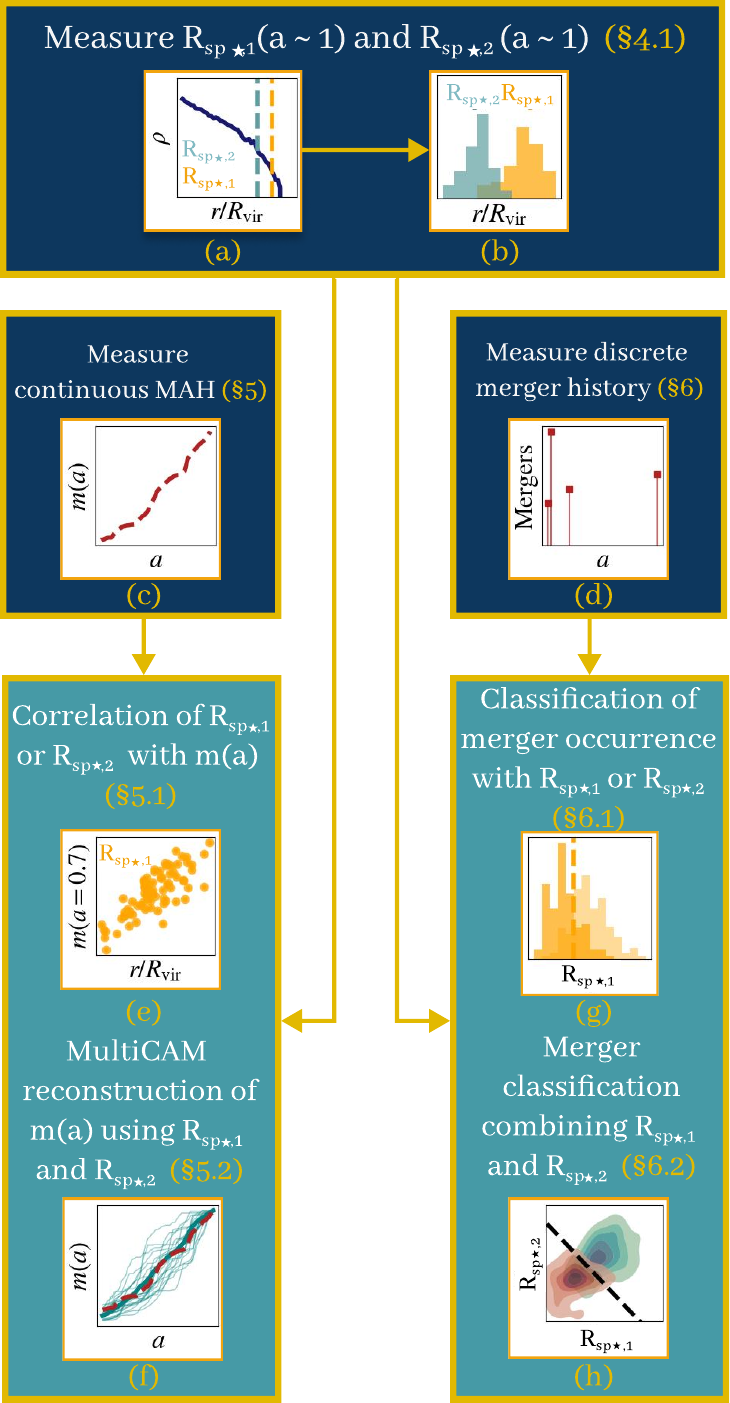}
    \caption{An illustration of the procedure of measurements and analysis as described in detail in \autoref{sec:Measure}, \autoref{sec:continuous}, and \autoref{sec:discrete}. We measure the present-day primary and secondary stellar splashback radii of simulated clusters (see \autoref{sec:Measure}) and probe how they inform the continuous mass accretion histories (MAH) (\autoref{sec:continuous}) and discrete merger histories of the clusters (\autoref{sec:discrete}). See the text in \autoref{sec:structure} for a description of each panel and toy figure.}
    \label{fig:flowchart}
\end{figure}

\begin{figure*}[t]
\includegraphics[width=\textwidth]{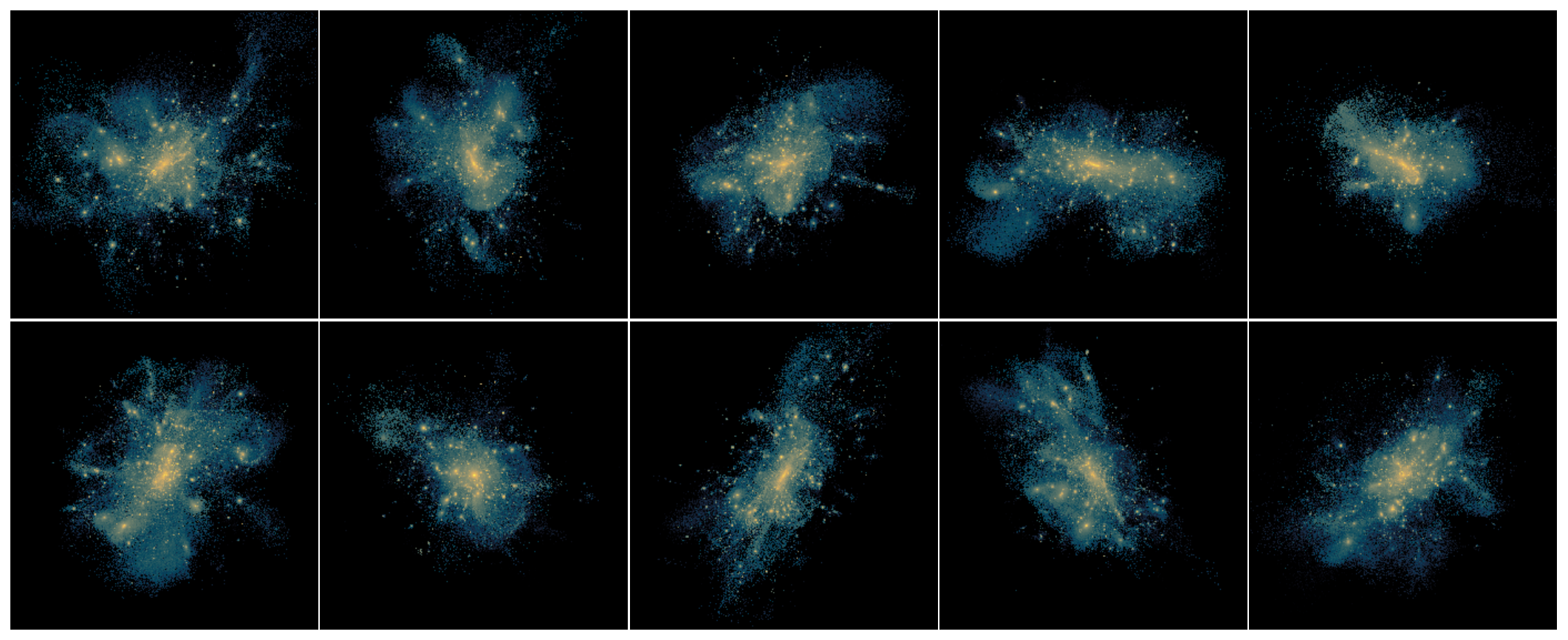}
    \caption{Projected stellar density maps for ten example clusters out of our sample of 79 galaxy clusters illustrating the diversity of stellar distributions captured by the \textsc{Nimbus} star-tagging framework.}
    \label{fig:visuals}
\end{figure*}

\section{Structural Overview}
\label{sec:structure}

The structure of this work is as follows. In \autoref{sec:sim}, we describe the high-resolution simulations and star-tagging framework used along with our simulated cluster sample. We illustrate the process of our subsequent measurements and analysis in \autoref{fig:flowchart}. As shown in the upper panel of \autoref{fig:flowchart}, we first measure the stellar splashback radii (see \autoref{sec:Measure}). The toy figures show measured stellar splashback radii from the stellar mass density profile (a) resulting in distributions of primary and secondary stellar splashback radii for our sample of clusters (b). In the center left panel, we measure the continuous MAH for each cluster with the toy figure (c) illustrating the mass of a cluster as a function of cosmological scale factor (see \autoref{sec:continuous}). In the lower-left panel, we correlate the stellar splashback radii with the continuous MAH as shown by the toy scatter plot (e) of the primary stellar splashback plotted against the mass of the cluster at a specific scale factor (see \autoref{subsec:correlation}). We then depict the reconstruction of sample continuous MAHs for each cluster given a pair of stellar splashback radii in toy figure (f) (see \autoref{subsec:multicam}). In the center-right panel, we measure the discrete merger history of each cluster with toy figure (d) illustrating discrete merger events with different merger ratios occurring at a range of scale factors (see \autoref{sec:discrete}). In the lower right panel, we classify the occurrence of mergers based on the values of each stellar splashback radii as illustrated in toy figure (g) by a distribution of stellar splashback radii corresponding to clusters that experienced a past merger (darker yellow) and a distribution that did not (lighter yellow) (see \autoref{subsec:1D}). We also illustrate in toy figure (h) the same exercise but classifying in the multivariable space of both stellar splashback radii (see \autoref{subsec:2D}).
In \autoref{sec:obs}, we discuss the range of surface brightnesses potentially occupied by these faint outer features of the ICL and the potential for observing them with next-generation imaging surveys. In \autoref{sec:conclusions}, we summarize and conclude.

\section{Simulations and star-tagging framework}
\label{sec:sim} 

In this work, we leverage the Symphony simulation suite \citep{Nadler_Mansfield_Wang_Du_Adhikari_Banerjee_Benson_Darragh-Ford_Mao_Wagner-Carena_et}, a compilation of cosmological cold dark matter (CDM) zoom-in simulations spanning nearly four orders of magnitude in host halo mass. Here, we focus on the most massive suite, \textsc{SymphonyCluster} \citep{wu2013rhapsodyi, wu2013rhapsodyii}, which enables us to provide a controlled sample of massive clusters with well-resolved outskirts. These hosts were selected from the Carmen simulation \citep{McBride2009} and have $z = 0$ virial masses between $8.2 \times 10^{14} \, M_{\odot}$ and $1.0 \times 10^{15} \, M_{\odot}$, with a mean of $9.2 \times 10^{14} \, M_{\odot}$.  The simulations adopt a flat $\Lambda$CDM cosmology with parameters $h = 0.7$, $\Omega_{\rm m} = 0.25$, $\Omega_{\Lambda} = 0.75$, $\sigma_{8} = 0.8$, and $n_s = 0.96$. The dark matter particle mass is $1.8 \times 10^8 \, M_{\odot}$, and the comoving Plummer-equivalent gravitational softening length is $3250 \, h^{-1} \, \mathrm{pc}$. Each simulation contains 200 snapshots, logarithmically spaced between scale factors $a = 0.075$ and $a = 1.00$, corresponding to approximately 25 snapshots per dynamical (crossing) time at $z = 0$.

We identify halos and subhalos using the \textsc{Rockstar} halo finder \citep{Behroozi2013_Rockstar} and construct merger trees with \textsc{Consistent-Trees} \citep{Behroozi2013_ConsistentTrees}.

\subsection{Star-tagging}
\label{subsec:star_tagging}

To model the distribution of stars in the ICL, we apply the \textsc{Nimbus} star-tagging framework (Mansfield et al., in prep). \textsc{Nimbus} assigns stellar mass to dark matter particles in infalling subhalos based on orbital energy and empirical structural relations, allowing us to model the disruption and deposition of stellar mass into the ICL in a realistic and dynamically stable manner.
This method builds on a well-established body of work that typically involves tagging the most-bound fraction of dark matter particles with stellar mass \citep[e.g.,][]{DeLuciaHelmi2008, Cooper2010, Bailin2014, Deason2022_dwarfs}, and in some cases, performing abundance matching on particle energy distributions \citep{BullockJohnston2005}. To produce more realistic orbital structures, \textsc{Nimbus} constructs a galaxy’s stellar density profile by superposing subsets of particles with time-stable radial distributions. This approach enables the generation of stellar halos that are dynamically stable and can flexibly match a wide range of target density profiles.
We illustrate the diversity of cluster stellar distributions star-tagged with \textsc{Nimbus} in \autoref{fig:visuals}. For a more detailed description of the \textsc{Nimbus} methodology and its limitations, we refer the reader to Mansfield et al., in prep. As we argue in Appendix \ref{app:Rhalf}, even substantial changes in the size of our tagged galaxies or meaningful changes in the shape of the stellar surface density profiles have negligible impacts on the location of the splashback radius, so it is unlikely that the details of our star-tagging scheme strongly impact our results.

We outline the specific relations that constitute our galaxy--halo connection model. We use the \textsc{UniverseMachine} stellar mass--halo mass (SMHM) relation fit from \citet{Behroozi2019} with a scatter of 0.2 dex \citep{Wechsler_Tinker_2018}. We use the size--virial radius relation from \citet{Jiang2019} that fits galaxy sizes as a function of redshift and concentration using the redshift-dependent abundance matching results of \citet{Somerville2018} based on GAMA and CANDELS survey galaxies. This relation is as follows:
\begin{equation}
\label{eq:Jiang}
\frac{r_{\star}}{r_{\rm vir}} = 0.02 (1+z)^{-0.2} (c/10)^{-0.7}
\end{equation}
where $r_{\star}$ is the 3D stellar half-mass radius and $c$ is the concentration. We include a 0.2 dex scatter in this relation.  In \autoref{app:Rhalf}, we examine the impact of drastically varying the size--virial radius relation on the measured stellar splashback radii and demonstrate that they are only slightly sensitive to such extreme variations in the relation.
The satellite galaxy stellar density profiles are parameterized by de-projected S\'ersic profiles, a default option in \textsc{Nimbus} that determines a profile shape given $r_{\star}$, the 3D stellar half-mass radius obtained with the \citet{Jiang2019} relation, and $M_{\star}$, the stellar mass using the above SMHM relation. The S\'ersic index $n_{\rm Sersic}$ of the S\'ersic profile is determined using an $M_{\star}-n_{\rm Sersic}$ relation fit to a large stellar mass range of galaxies from the GAMA survey as well as lower mass satellite galaxies around the Milky Way and M31. A detailed discussion of the de-projected S\'ersic profiles and S\'ersic indices can be found in Mansfield et al., in prep. We test the impact of $n_{\rm Sersic}$ on our ICL measurements by fixing the S\'ersic index for all satellite galaxies and varying over $0 \leq \textrm{log}_{10}\left(n_{\rm Sersic}\right) \leq 0.75$ and find that our measurements of the stellar splashback radii vary by less than 5\%. We also note that using an alternate profile such as the Plummer model does not alter our conclusions.    

We note potential caveats for this star-tagging scheme in the context of the ICL. The host dark matter halo does not get tagged with stars and, therefore, there is neither a brightest central galaxy (BCG) nor an ``in-situ" component to our ICL. Though the contribution of in-situ stars to the ICL is widely considered to be subdominant to other assembly channels, the exact fraction and its radial dependence is an ongoing topic of research.
Observational studies of the inner 100 kpc of individual clusters' ICL have estimated in-situ fractions as low as 1\% \citep{Melnick2012} or as high as 15-21\% \citep{Barfety2022}. In simulations, \citet{Puchwein2010} found a $\sim$30\% in-situ ICL fraction, but acknowledged that their outer halo star formation could be spurious.
More recent IllustrisTNG simulation studies of similarly massive clusters to our sample have shown in-situ ICL fractions of less than  $\sim$5-10\% within radial ranges of 100 kpc and $R_{200\rm c}$ \citep{Pillepich_Nelson_Hernquist_Springe_RüdigerPakmor_Torrey_Weinberger_Gene_Naiman_Marinacci_etal._2018} and 30 kpc and $R_{500\rm c}$ \citep{Montenegro2024}, though see \citet{Ahvazi2024} for a recent study claiming diffuse star formation-driven in-situ ICL fractions of $\sim$8-28\% in three lower mass TNG50 clusters. The Horizon Run 5 and Horizon-AGN simulations have also recently both found in-situ ICL fractions of less than $\sim$0.5\% for host halos at slightly lower mass than our sample \citep{Brown2024, Joo2024}. 

We stress that the ICL features we probe in this work are beyond 1 Mpc from the cluster center and therefore unlikely to host any significant in-situ stellar component from processes in a cluster's core. We also note that observationally, it is nontrivial to separate the BCG from the ICL of a cluster and it has been suggested that the two should be treated as one system (see \citealt{Montes_2022} and references therein). However, the ICL features beyond 1 Mpc should also not be affected by this transition from the ICL to the BCG, as it occurs at significantly smaller clustocentric distances.
In this analysis, we purely consider the ex-situ ICL built up by stripped and merging galaxies (i.e., star-tagged subhalos) and claim that any small in-situ component or BCG contribution extending to such large clustocentric distances could be considered an as yet unmodeled observational effect on our theoretical measurement of features of the ex-situ ICL. 
 
Two other caveats to this star-tagging method are the lack of a self-gravitating stellar component and the lack of central disk potentials that can impact the satellite galaxies that infall onto the cluster. The former might cause satellite galaxies to disrupt more readily than if their tagged stars increased their bound mass and the latter might effect the tidal stripping and disruption of satellites \citep[e.g.][]{Garrison-Kimmel2017, Wang2024}, thereby affecting the buildup of the ICL. We leave formal analyses of the impact of these missing components to future work.

\begin{figure*}[t]
\includegraphics[width=\textwidth]{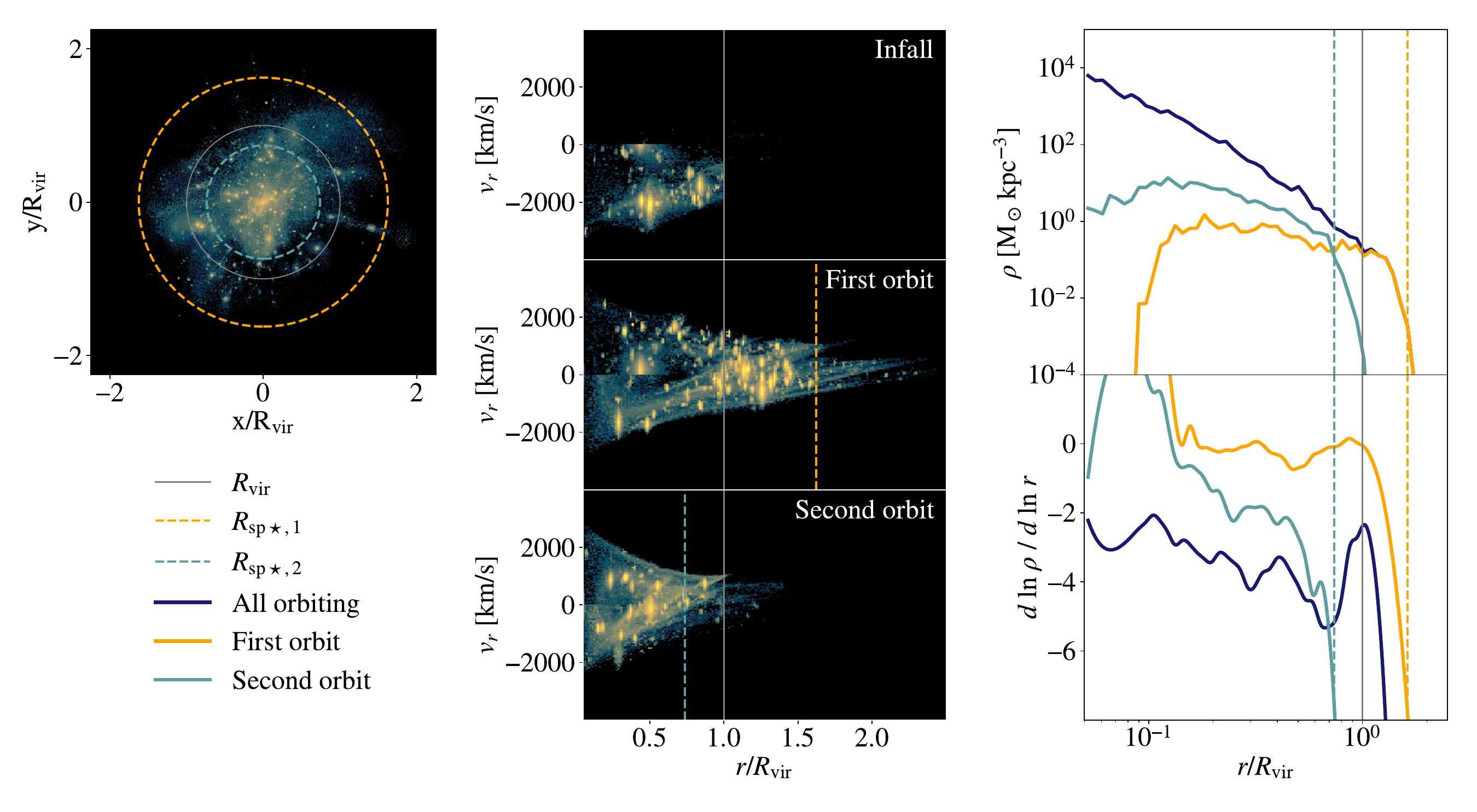}
    \caption{An example of the stellar density maps, phase space distributions, and radial stellar mass density profiles and our separation of particle populations for a cluster following the methods described in \autoref{subsec:MeasureRsp}. In the left column, we show the projected stellar mass density map of the cluster and indicate the relevant radii. In the central column, we show the density of stellar mass in radial velocity phase space separated into particles on infall, first orbit, and second orbit. As particles are only star-tagged after crossing $R_{\rm vir}$, the infall panel is lacking any stars outside that radius. In the right column, we show the profiles and corresponding log derivatives of particles on all orbits, on first orbit, and on second orbit. The dashed circles and vertical lines indicate the resulting measured primary and secondary stellar splashback radii for this cluster.}
    \label{fig:phaseprof}
\end{figure*}

\section{Measurement of present-day cluster properties}
\label{sec:Measure}
We aim to probe the historical information imprinted in the present-day properties of the outer ICL and compare how much information about previous mass growth is contained in these outer stellar features compared to more traditional tracers of formation history.  
In this section, we summarize the simulated cluster properties studied in this paper and our measurement methods.

\subsection{Measurement of Stellar Splashback Radii}
\label{subsec:MeasureRsp}
We center our analysis on two idealized measurements of the edges of the stellar halo, which we refer to as the primary and secondary stellar splashback radii. These measurements described in this section correspond to the toy depictions presented in the upper most box of \autoref{fig:flowchart}.

The term ``splashback radius" commonly refers to a physically motivated boundary to dark matter halos that separates infalling matter from orbiting matter at the apocenter of particles on their first orbit.  
In idealized spherical systems, the splashback feature is a sharp density caustic created by the pileup of particles with zero radial velocity at the apocenters of their orbits.
In the radial density profiles of a non-idealized and triaxial halo, this feature appears as a steepening in the density profile slope near this radius.
Thus, the splashback radius is often measured as the minimum of the log derivative of the radial density profile \citep{DK14, Adhikari2014, More2015, Shi2016}. 

This operational definition of the splashback radius is often applied to stacks of radial density profiles as opposed to individual halo profiles in simulations and observations \citep{DK14, Adhikari2014, More2015, Adhikari2016, More2016}. Stacking radial density profiles reduces the noise from individual cluster profiles but can also both smear out splashback radii at different radii and bias measured splashback features due to the presence of massive subhalos (see Section 4.2 in \citealt{Mansfield_2017} for a detailed analysis of this latter effect). Besides density profile methods, the splashback features in individual simulated halos have been accurately located using algorithms that leverage the three-dimensional density field \citep{Mansfield_2017} or particle orbits \citep{Diemer_2017}. We aim to relate the features of individual clusters' ICL to their mass accretion histories and therefore focus on idealized measurements for individual clusters as opposed to stacks.

The splashback radius typically refers to the feature made by particles on their first orbit. For clarity, we hereafter refer to this feature as the primary splashback radius or $R_{\rm sp, 1}$. Particles that have orbited twice will imprint the edge of their orbital distribution at their apocenter as a secondary slope steepening on the density profile. Sometimes referred to as the ``second caustic", this secondary splashback radius or $R_{\rm sp, 2}$ appears at smaller radii compared to the primary splashback radius. This can be understood as a consequence of the characteristic size of a particle's orbit scaling with the virial radius $R_{\rm vir}$ at the time of its infall (see Figure 2 in \citealt{Diemer_2017} for a depiction of a similar scaling with $R_{\rm 200m}$). As $R_{\rm vir}$ grows with time, so do the sizes of particles' orbits, resulting in larger apocenter radii for particles on their first orbit as compared to those on their second. Though there are smaller-radius, higher-order splashback radii which correspond to every integer number of orbits, we restrict ourselves to only the first and second splashback radii in this work.

The traditional splashback radius discussed above is defined in the dark matter distribution. We will refer to the analogous splashback features imprinted on the ICL as the primary and secondary ``stellar" splashback radii or $R_{\rm sp\star, 1}$ and $R_{\rm sp \star, 2}$, respectively. 
The stellar splashback radii are formed in a similar way to the dark matter splashback radii, except that instead of forming from all the halo's accreted matter, they consist of star particles which have been tidally stripped from accreted satellites.

Star particles in the simulation are only tracked when they have crossed $R_{\rm vir}$ of a host halo, leading to a potential artificial buildup of particles within $R_{\rm vir}$. As we aim to study the splashback effect that manifests as a steepening of the outer density profile, we exclude the infalling population to avoid artificially enhanced density at $R_{\rm vir}$
However, while steepening of the profile occurs at $R_{\rm sp\star, 1}$, there is no adjacent shallower profile made by infalling particles to form a minima. Therefore, instead of measuring the stellar splashback radii as minima of the log density profiles, we identify them using both density profiles and individual particle orbits.
We define $R_{\rm sp\star, 1}$ and $R_{\rm sp \star, 2}$ to be the radius corresponding to a fixed cutoff in the log derivative (in other words, a cutoff in the slope) of the density profiles of star particles on their first and second orbits, respectively.

We define particles on their first orbit to be those that have passed the first pericenter of their orbit but not the second. We similarly define particles on their second orbit to be those that have passed their second pericenter but not their third. We therefore begin by measuring the first, second, and third pericenter of each particle that falls into each cluster. In a simplification of the method used in \citet{Diemer_2022}, we define a given particle's pericenter passage as its crossing from negative to positive radial velocity, restricted to when the particle is within $1 R_{\rm vir}$ of the host cluster. Our determination of pericenter requires two snapshots with negative radial velocity followed by two snapshots with positive radial velocity. The third snapshot of those four is then labeled as that particle's first snapshot post-pericenter. We conducted a review of 300 particle orbits at a range of snapshots and judged whether the above pericenter-finding algorithm qualitatively matched our assessment of pericenter time based on manual inspection.
This test showed that our pericenter-finding algorithm could identify first pericenter correctly $\sim$98\% of the time, second pericenter over $\sim$95\% of the time, and third pericenter over $\sim$90\% of the time. Much of the decreasing accuracy is just caused by the output cadence of the simulation. Since our snapshots are logarithmically spaced in $a(t)$, second and third orbits will always be less well-sampled than first orbits. We also find that most of our our inaccuracy is limited to pericenters at early times; snapshots at later times still achieve $\sim$98\% accuracy for third pericenter. In our analyses of particles on their first and second orbits, we consider only particles that have undergone infall after the fifth snapshot ($a=0.08$) in order to mitigate the effects of early pericenter misidentification of particles that sank to the center of the cluster early on and remained there. This does not impact our results as we only analyze particles on their first and second orbits at late times. In the central panels of \autoref{fig:phaseprof}, we show the star particles of an example cluster separated into those on infall, first orbit, and second orbit in radial velocity phase space.

In order to measure $R_{\rm sp\star, 1}$ and $R_{\rm sp \star, 2}$ for each cluster, we then measure the radial stellar mass density profiles of the particles determined to be on on their first and second orbits, respectively. Instead of the commonly used spherically-averaged radial profile, we use an angular median radial profile to mitigate the impact of massive subhalos on the more general shape and features of the stellar profile (see Sections 4.2 and 4.3 of \citealt{Mansfield_2017} for an analysis of the effects of massive subhalos on the radial density profile and splashback measurements). The angular median profile method \citep{Mansfield_2017} consists of dividing each spherical shell into a fixed number of solid-angle segments and taking the median mass density value of the segments to construct the radial profile. We use 50 logarithmically distributed radial bins between 0.05 Mpc and 2.5 Mpc ($\sim30$ per decade) and 98 solid-angle segments for each radial shell chosen with a two-hemisphere variation on the algorithm presented in \citet{gringorten1992}.  In the upper right panel of \autoref{fig:phaseprof}, we show the angular median method radial stellar mass density profiles of all particles and those on their first and second orbits for an example cluster. 

We note that this angular median method, also employed by \citet{Deason_StellarSplashback}, is distinct to observational methods of ICL measurement, which include surface brightness cuts, composite model fitting, and increasingly complex algorithms for isolating the diffuse light (see \citealt{Montes_2022} and references therein). It is also distinct to the use of only unbound stars in simulated ICL or stellar halo measurements. However, our method allows us to probe the underlying diffuse distribution of ICL stars without making nontrivial decisions about which particles are bound or unbound to substructure. While it might not be a straightforward comparison between the angular median ICL profile in simulations and observational ICL profiles, this angular median method of studying underlying features of the ICL in simulations and observations is potentially less impacted by small definitional changes of the ICL. We stress that this initial exploration of the information held in features of the ICL profile is highly idealized.

Following our measurement of the radial density profiles, we use a fourth-order Savitsky--Golay algorithm \citep{Savitzky-Golay} to smoothly obtain the log derivatives of the profiles using a window size of 15 radial bins. In \autoref{app:spdef_tests}, we demonstrate that one of our key results, the correlation of $R_{\rm sp \star, 1}$ with the MAH, is trends only slightly with this choice of window size. As we only include particles on certain orbits around the host and not those infalling, the slope of the log derivative will continue falling to negative infinity. Therefore, we pin our measure of the stellar splashback radius to a slope cutoff of $-8$, which roughly aligns with the minimum slope in the total profile that would indicate $R_{\rm sp \star, 2}$. The exact slope cutoff chosen is not critical as we aim to study the relation between the ``edge" of the profiles and the assembly histories of the cluster and consistently apply this definition of the edge over all the clusters. Nevertheless, in \autoref{app:spdef_tests}, we find that our results are largely insensitive to this choice of slope cutoff. Because of this insensitivity, we do not fine-tune our window size or slope cutoff to maximize the strength of our results and instead continue with our initial choices: the ones stated in this Section. The radii corresponding to this slope cutoff for each particle population of each cluster results in idealized $R_{\rm sp\star, 1}$ and $R_{\rm sp \star, 2}$ measurements for the cluster sample. We make this measurement at a range of timescales. In this work, we focus on our ``present-day" snapshot, set to be scale factor $a = 0.99$. We choose this snapshot in order to have the full pericenter information, which requires a snapshot after the one of interest ($a = 1.0$) as described above. In the lower right panel of \autoref{fig:phaseprof}, we show the log derivatives of the profiles and the resulting measured locations of $R_{\rm sp\star, 1}$ and $R_{\rm sp \star, 2}$ for an example cluster at $a = 0.99$.

\subsection{Measurement of Other Present-day Properties}
\label{subsec:MeasureOther}

We contextualize the information contained in the stellar splashback radii by comparing to a few commonly used tracers of formation history and relaxedness. In this section, we describe the measurement of these properties in the simulation. 

One halo property we consider is $x_{\rm off}$, the offset between the center of mass of the dark matter halo and its most bound particle, normalized by the virial radius of the halo:
\begin{equation}
    x_{\rm off} = X_{\rm off}/R_{\rm vir}.
\end{equation}
This spatial offset is considered to be a tracer of the dynamical state or ``relaxation" of a cluster \citep{PowerKnebeKnollman2012} as well as correlated to age indicators \citep{Neto2007}. The offset and its relation to mass growth can be intuitively understood as a result of recent mergers disturbing the central regions of a cluster and creating deviations from spherical symmetry. Observationally, this offset can be probed by the separation between the BCG and the center of the dark matter distribution as probed by weak lensing or the center of the hot gas distribution as probed by X-ray observations \citep{Yoo2024}. 

We also consider the concentration $c_{\rm vir} = R_{\rm vir}/R_{\rm s}$, where $R_{\rm s}$ is the scale radius in an NFW profile. Instead of fitting an NFW profile to determine $R_{\rm s}$, $c_{\rm vir}$ is calculated using the maximum circular velocity $v_{\rm max}$, the virial velocity $V_{\rm vir} = \sqrt{\frac{G M_{\rm vir}}{R_{\rm vir}}}$, and NFW assumptions \citep[e.g.,][]{klypin2011}. This is done by solving the following equations for $c_{\rm vir}$:

\begin{align}
    \frac{c_{\rm vir}}{f(c_{\rm vir})} = \frac{V^2_{\rm max}}{V^2_{\rm vir}} \frac{x_{\rm max}}{f(x_{\rm max})} \\
    f(x) \equiv \text{ln}(1+x) - \frac{x}{1+x},
\end{align}
where $x_{\rm max} \equiv R_{\rm max}/R_s \approx 2.1626$. Halo concentration has long been connected with the mass assembly histories of halos. Intuitively, the central densities of halos grow rapidly in response to high mass accretion that reaches inner regions and slows as low mass accretion builds up accreted matter outside those regions. The significant impact of formation histories and mergers on halo concentrations has been examined in various works \citep[e.g.,][]{Wechsler2002,Zhao2003, Ludlow2012, Rey2019, KWang, Johnson2021, Stevanovich2023}.   

The final property we compare to is $M_{\star \rm 1,2}$, the stellar mass gap. This property is an analog of the magnitude gap, the magnitude difference between the two brightest galaxies in a cluster. Earlier forming halos are thought to have old central galaxies that have grown from mergers and are therefore substantially brighter than surrounding satellite galaxies, resulting in a noticeable magnitude gap. We define the stellar mass gap for each cluster to be the logarithmic difference between the highest and second highest stellar mass galaxy in the cluster:
\begin{equation}
    M_{\star \rm 1,2} = \text{log}(M_{\star \rm 1}/M_{\star \rm 2}).
\end{equation}
We use the SMHM relation including 0.2 dex of scatter as discussed in \autoref{subsec:star_tagging}. In order to account for the scatter in this SMHM relation, we use 1000 realizations of the stellar mass gap for each cluster and treat each realization as if it belongs to a separate cluster for our later analyses. We treat the stellar mass gap in this way as it is necessarily highly impacted by realizations of the stellar mass and the scatter in the relation. In essence, we study a probability distribution of the stellar mass gap. We further test the stellar mass gap $M_{\star \rm 1,4}$ defined using the first and fourth highest stellar mass galaxy in the cluster and find it to be less informative than $M_{\star \rm 1,2}$. The magnitude gap and the stellar mass gap have been examined as strong predictors for the formation time \citep[e.g.,][]{Jones2003,D’onghia_Sommer-Larsen_Romeo_Burkert_Pedersen_Portinari_Rasmussen_2005,vonBendaBeckmann2008, Dariush_Raychaudhury_Ponman_Khosroshahi_Benson_Bower_Pearce_2010, Deason2013, Raouf2014, Vitorelli2018} and dynamical state of clusters \citep[e.g.,][]{Zarattini2015, Raouf2016, Yoo2024, Casas2024, GoldenMarx2024_maggap}. 

\section{Continuous mass accretion history}
\label{sec:continuous}

Motivated by the relation between the dark matter splashback radius of clusters and their accretion rate, along with the ICL's role as a visible tracer of the dark matter distribution, we investigate how our idealized measurements of stellar splashback radii inform the continuous MAHs of the massive clusters in the Symphony simulation suite. This section of our analysis corresponds to the left-hand panels of \autoref{fig:flowchart}. As illustrated in the center left panel of \autoref{fig:flowchart}, we measure the continuous MAH as the mass of the cluster as a function of time or scale factor. In \autoref{sec:discrete}, we contrast this with the discrete merger history, the major mergers experienced by a cluster over time. In \autoref{subsec:correlation}, we find that the present-day stellar splashback radii are strongly correlated with the continuous MAHs of our cluster sample, particularly at certain timescales. In \autoref{subsec:multicam}, we use the stellar splashback radii to reconstruct the MAHs of individual clusters.

We first present our notation and parametrization of the continuous MAH. We measure time using the cosmological scale factor as well as in units of dynamical time. The cosmological scale factor is defined in terms of redshift $z$ as:
\begin{equation}
    \label{eq:a}
    a(z) = \frac{1}{1+z}.
\end{equation}
The dynamical time associated with halos at each redshift $z$ can be defined as the time it takes to cross 2$R_{\rm vir}$ at that given $z$:
\begin{equation}
\label{eq:tdyn}
    t_{\rm dyn}(z) = \frac{2R_{\rm vir}}{V_{\rm vir}} = \frac{2R_{\rm vir}}{\sqrt{\frac{G M_{\rm vir}}{R_{\rm vir}}}} = \sqrt{\frac{3}{\pi G \rho_{\rm vir}(z)}},
\end{equation}
where $V_{\rm vir}$ is defined as a typical velocity at $R_{\rm vir}$ and $\rho_{\rm vir}$ is the enclosed density within $R_{\rm vir}$ as calculated with the \citet{BrynNorman1998} convention. Note that all halos at a given redshift have the same $\rho_{\rm vir}$. Therefore, $t_{\rm dyn}$ depends only on cosmology and redshift, not on the specifics of halos. Following \citet{Jiang_van_den_Bosch_2016} and \citet{KWang}, we define $T$ as a measure of time in units of dynamical time:
\begin{equation}
    \label{eq:T}
    T(a ; a_{\rm ref}) = \int^{t(a)}_{t(a_{\rm ref})} \frac{dt}{t_{\rm dyn}(t)},
\end{equation}
where $t(a)$ is the age of the universe at a given scale factor $a$ and $a_{\rm ref}$ is the reference scale epoch. 
Following the parametrization of \citet{Mendoza}, we measure monotonic MAHs for each cluster using the peak virial mass normalized with the $a = 0.99$ peak virial mass:
\begin{equation}
\label{eq:ma}
    m(a) = \frac{M_{\rm peak}(a)}{M_{\rm peak}(a = 0.99)},
\end{equation}
where $M_{\rm peak}$ is defined as

\begin{equation}
\label{eq:mpeak}
    M_{\rm peak}(a_{\rm i}) = \text{max}_{\rm j \le \rm i} [M_{\rm vir}(a_{\rm j})].
\end{equation}
This mass fraction relative to a cluster's present-day mass $m(a)$ can also be written as $m(T)$, a function of dynamical times from an $a = 0.99$ reference epoch, using \autoref{eq:T}.

\begin{figure}[t!]
\includegraphics[width=\columnwidth]{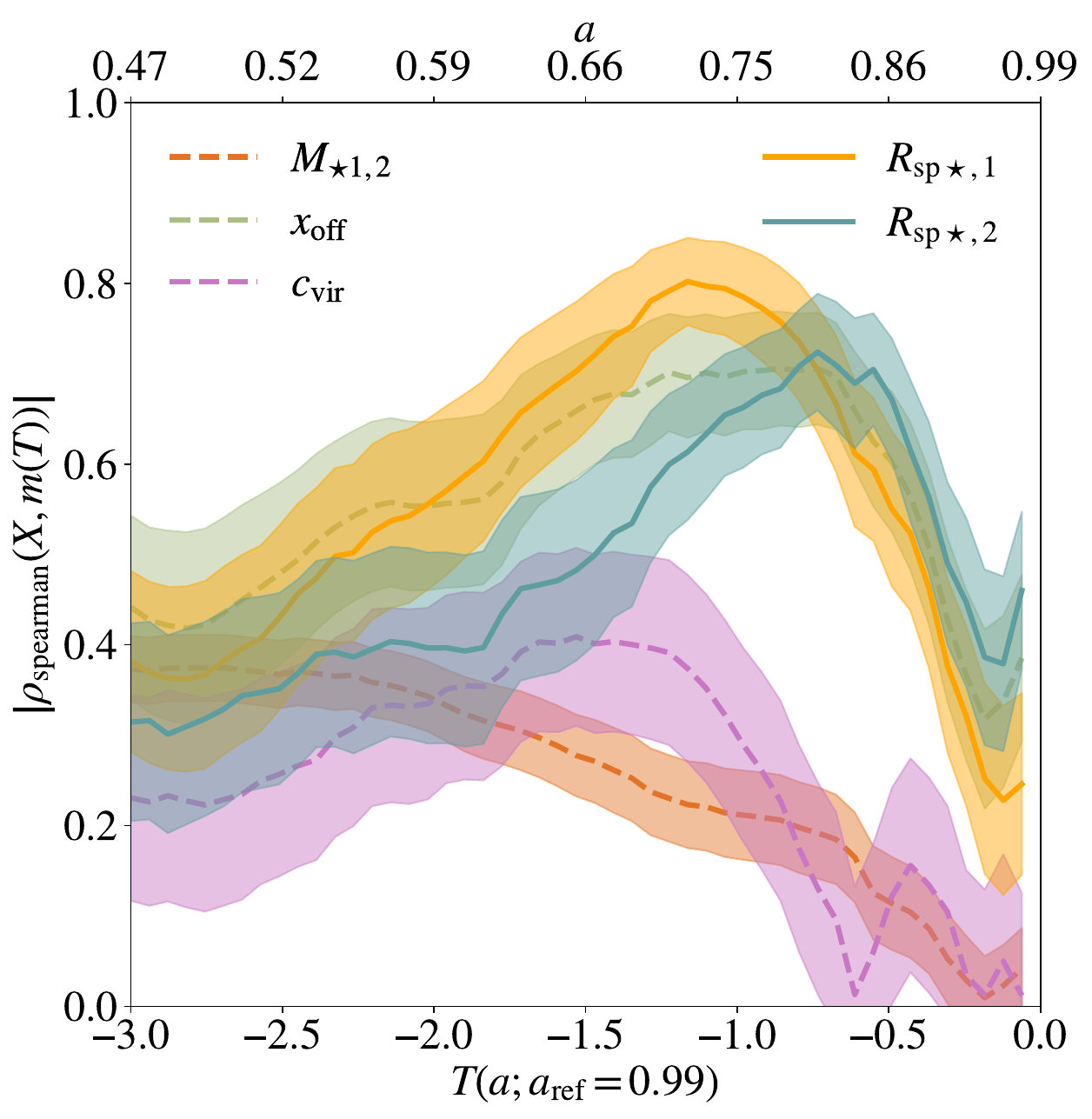}
    \caption{Spearman's correlation coefficients between present-day cluster properties and the mass accretion history (MAH) as a function of time. Time $T$ is measured in units of number of dynamical times (where $T$= 0 indicates present-day) and corresponding scale factors are included in the upper $x$-axis for reference. The shaded bands indicate the jackknife-estimated standard error (see \autoref{subsec:correlation} for further details). We plot our proposed tracers of the MAH (solid lines): the primary stellar splashback $R_{\rm sp\star, 1}$ and the secondary stellar splashback $R_{\rm sp \star, 2}$. Traditional tracers of the MAH are shown for comparison (dashed lines): the central offset $x_{\rm off}$, the concentration $c_{\rm vir}$, and the stellar mass gap $M_{\star \rm 1,2}$. $R_{\rm sp\star, 1}$ is highly correlated with $m(T)$ with a peak 1.16 dynamical times ago. $R_{\rm sp \star, 2}$ is strongly correlated to a slightly lesser degree and peaks more recently, 0.73 dynamical times ago. $x_{\rm off}$ is correlated similarly to $R_{\rm sp\star, 2}$, but  $M_{\star \rm 1,2}$ and $c_{\rm vir}$ are both substantially less correlated with MAH than the stellar splashback tracers.}
    \label{fig:correlation}
\end{figure}

\subsection{Correlations with Continuous Mass Accretion history}
\label{subsec:correlation}

To probe the information that our measured features of the ICL contain about the continuous mass growth of clusters, we begin by considering the correlations between their present-day properties and MAHs. This analysis corresponds to the toy illustration in the upper part of the lower left panel in \autoref{fig:flowchart}.
Using the methods detailed in \autoref{subsec:MeasureRsp}, we measure the primary and secondary stellar splashback features as well as several traditional cluster properties of the 79 galaxy clusters in our sample at $a = 0.99$. In \autoref{fig:correlation}, as a function of time $T$, we determine the Spearman's rank correlation coefficient between each property and the MAH parameterized as $m(T)$ (see \autoref{eq:ma}). We measure time $T$ in units of number of dynamical times (see \autoref{eq:T}) and include the corresponding scale factors in the upper $x$-axis. In addition to our proposed tracers of MAH, the primary and secondary splashback $R_{\rm sp\star, 1}$ and $R_{\rm sp \star, 2}$, we display correlation coefficients for traditional tracers: the normalized offset between the halo's center of mass and its most bound particle, $x_{\rm off}$, the stellar mass gap $M_{\star \rm 1,2}$, and the concentration $c_{\rm vir}$. The shaded bands around each curve indicate the jackknife standard error estimated using the leave-one-out method, rerunning the analysis removing each cluster one at a time.

\begin{figure*}[ht]
\includegraphics[width=\textwidth]{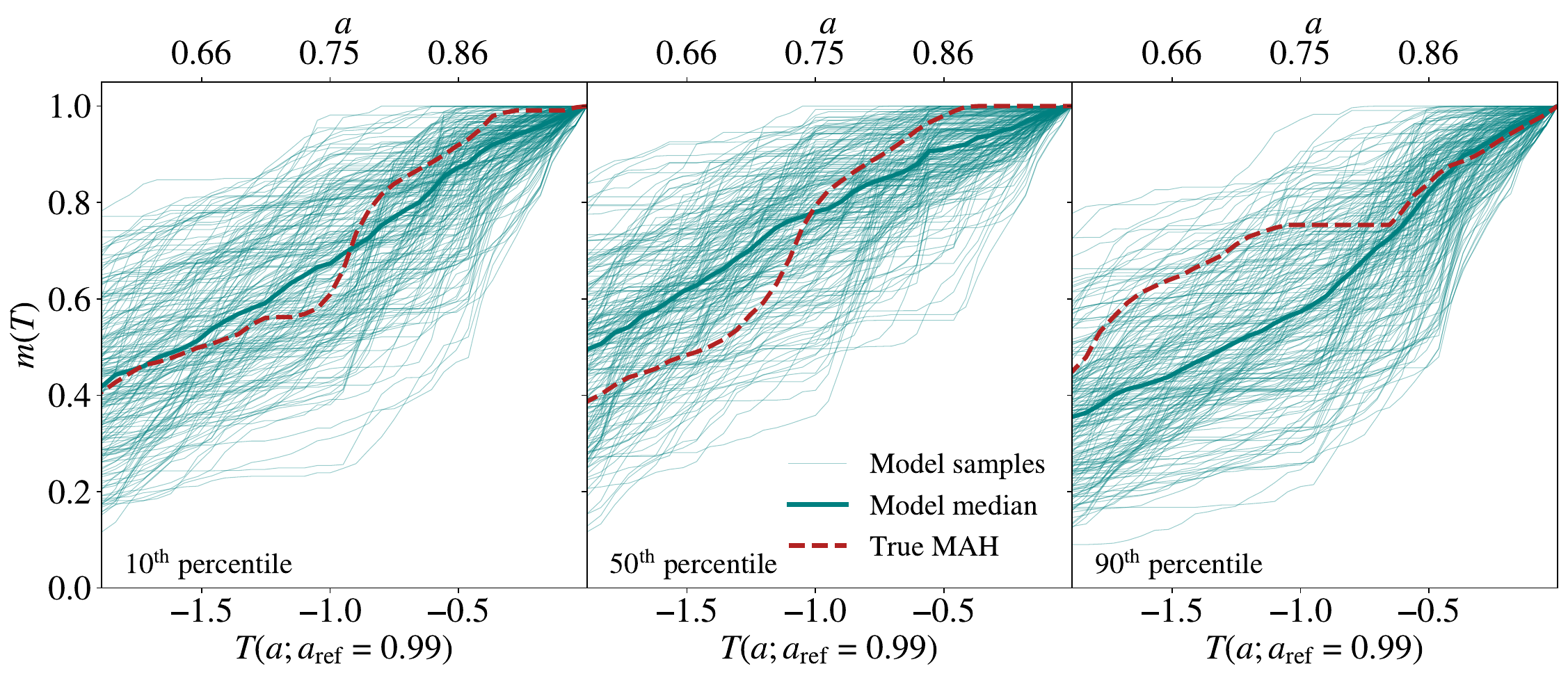}
    \caption{Example true MAHs, 200 \textsc{MultiCAM} model reconstructed MAH samples, and medians of \textsc{MultiCAM} model reconstructed MAH samples for three clusters. MAHs are plotted as $m(T)$ as a function of $T$, time in units of dynamical times where $T = 0$ is present day and the corresponding scale factors are indicating in the upper $x$-axis. The three clusters shown reflect three percentiles of accuracy in model predictions (where lower percentiles indicate higher accuracy). Accuracy is quantified here by the root mean squared error (RMSE) between the median reconstructed MAH and the true MAH. The 10$^{\rm th}$ percentile cluster has a closer recovery of true MAH in its median reconstructed MAH than the 90$^{\rm th}$ percentile cluster.}
    \label{fig:MAH}
\end{figure*}
The primary stellar splashback radius $R_{\rm sp\star, 1}$ is strongly correlated with the mass of the cluster, particularly between 2.0 and 0.5 dynamical times in the past (equivalently between $a$ $\approx$ 0.6 and $a$ $\approx$ 0.85), peaking at 1.16 dynamical times ago.
The secondary stellar splashback radius $R_{\rm sp \star, 2}$ is also strongly correlated, though to a slightly lesser degree, and peaks at a more recent timescale of 0.73 dynamical times ago. $R_{\rm sp \star, 2}$ has a secondary increase with lower correlation at $\sim$ 2.5 dynamical times ago. Given that $R_{\rm sp \star, 2}$ is the signature of particles on their second orbit, it would be natural to expect that it would probe earlier timescales instead of peaking at a more recent timescale than the primary stellar splashback. However, this difference of approximately half a dynamical time could be a result of infalling mass spending less time within the smaller radius of $R_{\rm sp \star, 2}$, leading to a shorter experience of and faster recovery from the gravitational contraction shrinking the splashback density features. This also aligns with the findings of \citet{Lucie-Smith_2022} that the outer profiles of halos are sensitive to several timescales including recent mergers. We also note that we find the correlation strengths of the stellar splashback radii to be largely insensitive to the inclusion of scatter in the SMHM relation. We also test how sensitive the correlation between $R_{\rm sp\star, 1}$ and a cluster's MAH is under variations in the stellar splashback measurement methodology in \autoref{app:spdef_tests}.   

\begin{figure*}[ht]
\centering
\includegraphics[width=\textwidth]{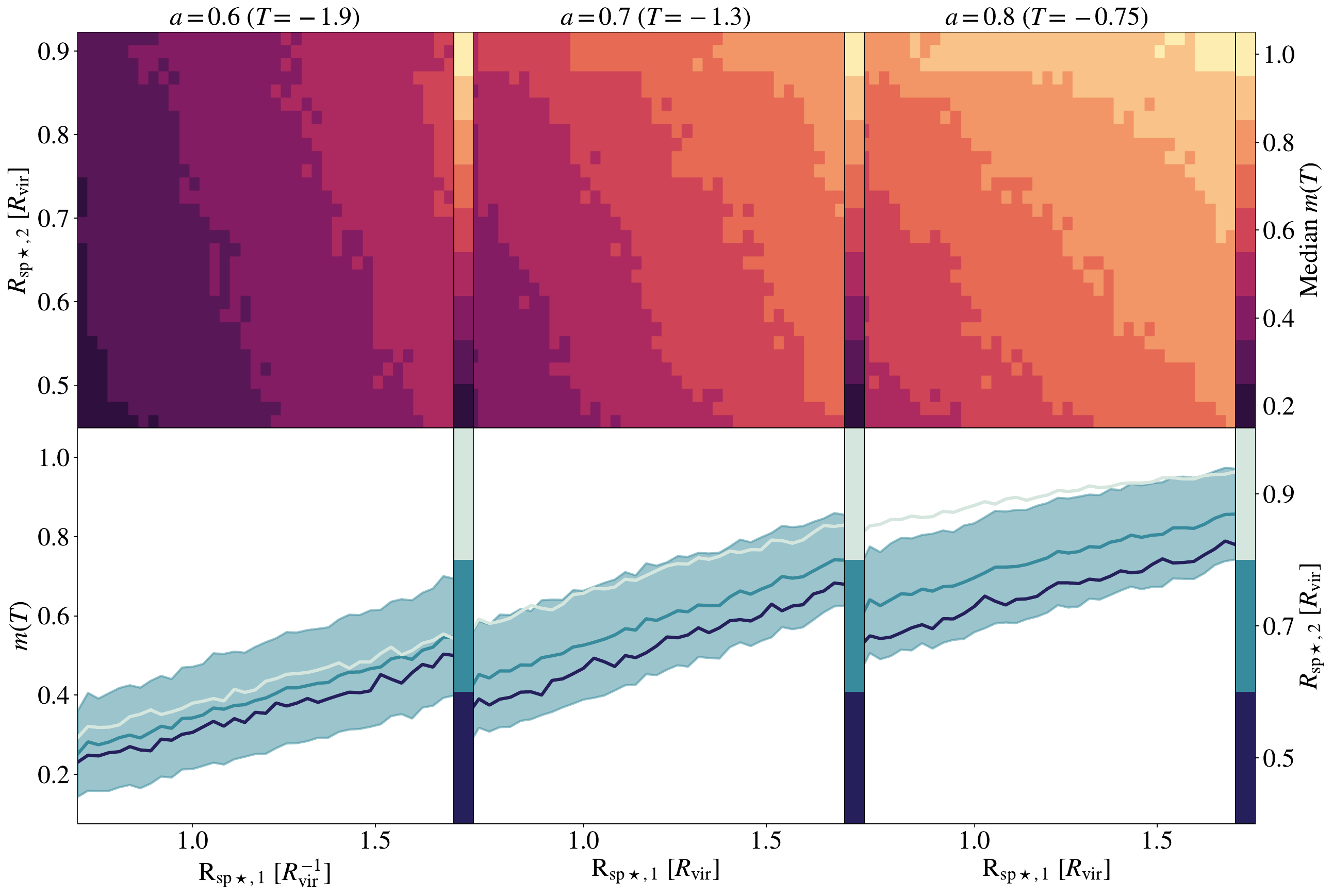}
    \caption{Predictions of the \textsc{MultiCAM} model for mass accretion history as a function of the primary and secondary stellar splashback radii, $R_{\rm sp\star, 1}$ and $R_{\rm sp \star, 2}$, created by generating 500 samples for each pair of $R_{\rm sp\star, 1}$ and $R_{\rm sp \star, 2}$. Upper panels: The median $m(T)$ of the 500 predicted samples generated by \textsc{MultiCAM} at three epochs (columns) as a function of the primary and secondary splashback, $R_{\rm sp\star, 1}$ and $R_{\rm sp \star, 2}$. Lower panels: The median $m(T)$ of the predicted mass accretion history samples at three epochs (columns) as a function of $R_{\rm sp\star, 1}$ for three selected $R_{\rm sp \star, 2}$ values. The RMSE scatter around the median for $R_{\rm sp \star, 2} = 0.7$ is shown as a shaded band around the curve. The lack of smoothness in the contours and the curves is a result of the stochasticity inherent in generating a finite number of samples. The sensitivity of the predictions to each stellar splashback radius varies depending on epoch, corresponding to the varying of correlation strengths in \autoref{fig:correlation}.}
    \label{fig:multicam_grid}
\end{figure*}
Comparing to our additional measured cluster properties, $x_{\rm off}$ has similar correlation strength to $R_{\rm sp\star, 2}$ and lower correlation strength than $R_{\rm sp\star, 1}$. The correlation curve of $x_{\rm off}$ is maximized at 0.86 dynamical times in the past, between the two stellar splashback radii, though it maintains similarly high correlation between $\sim1.2$ and $\sim0.7$ dynamical times ago. This is a particularly useful point of comparison as $x_{\rm off}$ is an idealized version of the traditional tracer of recent merger activity typically probed by the X-ray emission of hot gas or weak lensing mass distributions. The concentration $c_{\rm vir}$ is less correlated than both the stellar splashback radii and probes approximately 1.5 dynamical times ago. The stellar mass gap $M_{\star \rm 1,2}$ is correlated with MAH to a similar degree as $c_{\rm vir}$ and appears to probe earlier timescales with a broad peak at 2.69 dynamical times ago. As expected, the correlation strength of $M_{\star \rm 1,2}$ is sensitive to the inclusion of scatter in the SMHM relation. When the scatter is increased from 0.2 dex to 0.3 dex, we find that the peak of the correlation curve of $M_{\star \rm 1,2}$ decreases to $\rho_{\rm spearman}=0.2$.

\citet{Shin_Diemer_2022} performed a similar analysis of the dark matter splashback correlation with the continuous mass accretion history and found that the correlation of their high mass sample peaked $\sim1-1.5$ dynamical times ago. The consistency of our stellar splashback results with the dark matter splashback results of \cite{Shin_Diemer_2022} supports the use of these ICL features as visible tracers of dark matter dynamical signatures. Comparing to the analogous $x_{\rm off}$ Spearman's correlation curves shown in \citet{Mendoza} using the Bolshoi dark matter-only cosmological simulations, our $x_{\rm off}$ peak correlation epoch is at slightly more recent timescales and shows higher correlation strength. Our $c_{\rm vir}$ curve also peaks at more recent timescales and at lower $\rho_{\rm spearman}$ than for the halo sample of \citet{Mendoza}. That work included a large sample of significantly lower mass galaxies ($M_{\rm vir} \sim 10^{12} h^{-1}$ M$_{\odot}$). The differences in our results could potentially indicate a halo mass dependence in the timescales and strength of correlation with MAH.
The epoch of $c_{\rm vir}$'s peak correlation is in close agreement with the findings of \citet{KWang} for halos in a similar mass range. However, the amplitude of our correlation curve is lower than the $\rho_{\rm spearman}$ of 0.6 seen in their figure 1.

The high correlation between the two stellar splashback radii and MAH as well as the complementarity of the timescales probed open an intriguing avenue for reconstructing the MAH of an individual cluster using the outer features of its ICL distribution.

\subsection{Reconstruction of Continuous Mass Accretion History}
\label{subsec:multicam}
We have demonstrated that idealized measurements of the primary and secondary stellar splashback radii are highly correlated with the MAH.
The correlation curves shown in \autoref{fig:correlation} indicate that there are a range of timescales probed by the splashback radii and that notably, the two features probe different timescales.
In this section, we use the combined information of the two present-day stellar splashback radii to reconstruct the individual MAHs of our sample of clusters with an emphasis on the range of timescales best probed. This analysis corresponds to the toy illustration in the lower part of the lower left panel in \autoref{fig:flowchart}. We also show how the predictions of the model vary more generally as a function of $R_{\rm sp\star, 1}$ and $R_{\rm sp \star, 2}$ at a range of epochs.

To reconstruct the MAHs, we use the \textsc{MultiCAM} framework presented in \citet{Mendoza}. The \textsc{MultiCAM} algorithm incorporates a generalized version of conditional abundance matching to relate full MAHs of halos with single-epoch halo properties.
\textsc{MultiCAM} computes the Spearman cross-correlation matrix between all input quantities (halo properties) and output quantities (MAHs), generates samples from a simple multi-dimensional Gaussian with that cross-correlation matrix, and rescales the marginal distributions of that distribution to match those found in the training set. In this sense, it can be thought of as a very simple special case of a Gaussian Process. In cases where the distribution \textsc{MultiCAM} is modeling is well-described by a Gaussian copula, \text{MultiCAM} will generate exact samples from that distribution.

We provide \textsc{MultiCAM} the present-day $R_{\rm sp\star, 1}$ and $R_{\rm sp \star, 2}$ values and the MAHs for the corresponding clusters in our sample. To test \textsc{MultiCAM}'s predictions on reconstructing a given cluster's MAH, we leave that cluster out and train on the remaining 78 clusters. In order to enforce a monotonically increasing MAH, we choose to provide the differential MAH, which we define as:
\begin{equation}
\label{eq:dma}
    \delta m(a_{\rm i}) = \text{ln}[m(a_{\rm i+1})/m(a_{\rm i})],
\end{equation}
where $a_{\rm i}$ is the scale factor at a given snapshot as in \autoref{eq:mpeak}. Monotonicity is enforced in \textsc{MultiCAM}'s prediction by construction.
Since $m(a=1) = 1$ by definition (see \autoref{eq:ma}), the MAHs' $m(a)$ can be recovered from a given $\delta m(a)$. \textsc{MultiCAM} returns its predictions for given values of $R_{\rm sp\star, 1}$ and $R_{\rm sp \star, 2}$ in the form of a set of sample predictions that incorporate the scatter of the target distribution, as opposed to a single predicted MAH. We determine the accuracy of the predictions by comparing the median of the sample predictions to the true MAH in the range between $a = 0.6$ and $a = 0.9$ and calculate the root mean squared error (RMSE). We choose this range as our focus since this is the range of epochs we appear to have sensitivity to in \autoref{fig:correlation}. In \autoref{fig:MAH}, we plot several example clusters with their samples of predicted MAHs, the median of those samples, and their true MAHs. We choose example clusters to be at three percentiles (10$^{\rm th}$, 50$^{\rm th}$, 90$^{\rm th}$) in RMSE accuracy of the median of the predicted samples compared to the true MAH. Lower percentiles in RMSE indicate higher accuracy of the predictions. The 10$^{\rm th}$ percentile cluster has a closer recovery of true MAH in its median reconstructed sample MAH than the 50$^{\rm th}$ and 90$^{\rm th}$ percentiles clusters.

We also test the performance of \textsc{MultiCAM} using each stellar splashback radius individually. We find that the RMSE of the predicted MAHs can be $\sim 20\% - 40\%$ higher, particularly at earlier epochs, when only using $R_{\rm sp \star, 2}$ instead of both $R_{\rm sp\star, 1}$ and $R_{\rm sp \star, 2}$. The RMSE of predicted MAHs using $R_{\rm sp\star, 1}$ alone can be $\sim 20\% - 30\%$ higher at more recent epochs, reflecting the loss of information from $R_{\rm sp \star, 2}$ in that regime. 

We can also consider the predictions of the \textsc{MultiCAM} model more generally in the full space of $R_{\rm sp\star, 1}$ and $R_{\rm sp \star, 2}$, rather than limited to the specific stellar splashback radii and MAHs found in our sample of clusters. 
In \autoref{fig:multicam_grid}, for three epochs ($a = 0.6, 0.7, 0.8$), we show the predictions of the model in two ways. In the upper panels, we show the predictions in the two dimensional space of $R_{\rm sp\star, 1}$ and $R_{\rm sp \star, 2}$. For each point in $R_{\rm sp\star, 1}$ and $R_{\rm sp \star, 2}$ space, we generate 500 sample MAHs. The upper panels are colored by the median mass $m(T)$ of the predicted samples.
We limit the parameter space shown to the region in which we have more than one cluster on which \textsc{MultiCAM} can train. Note that, in principle, values of $R_{\rm sp\star, 1}$ and $R_{\rm sp \star, 2}$ can go beyond these imposed bounds. In the lower panels, we show the median mass $m(T)$ of the predicted samples as a function of $R_{\rm sp\star, 1}$ for three fixed values of $R_{\rm sp\star, 2}$. We include a shaded band around the $R_{\rm sp \star, 2} = 0.7$ curve indicating the scatter in \textsc{MultiCAM}'s predicted samples, quantified by the RMSE with respect to the median. This scatter is a conservative estimate \textsc{MultiCAM}'s accuracy in its predictions of the MAHs. We confirm that the scatter returned by the model is an upper bound on the RMSE of the model predictions in \autoref{app:scatter}. 

In the upper and lower panels of \autoref{fig:multicam_grid}, we can see that \textsc{MultiCAM} is reflecting the relation between larger stellar splashback radii and larger $m(T)$. This general trend reflects that later-forming systems with lower $m(T)$ followed by more recent and rapid mass accretion have smaller splashback radii due to gravitational contraction. Earlier forming systems with high $m(T)$ have gained little mass in recent times and have larger uncontracted splashback radii. Particularly visible in the upper panels, for $a = 0.6$, $m(T)$ varies strongly with $R_{\rm sp\star, 1}$, but less strongly with $R_{\rm sp\star, 2}$. However, at $a = 0.7-0.8$, $m(T)$ varies as strongly with $R_{\rm sp\star, 2}$ as with $R_{\rm sp\star, 1}$. This time evolution is reflecting the varying information content provided by $R_{\rm sp\star, 1}$ and $R_{\rm sp\star, 2}$ in the prediction of the MAH at different epochs. It exactly reproduces the trends in the Spearman's correlation coefficients between the present-day stellar splashback radii and $m(T)$ as seen in \autoref{fig:correlation}. $R_{\rm sp\star, 1}$ almost entirely dominates the correlation at $a = 0.6$ and even at $a = 0.7$, however $R_{\rm sp \star, 2}$ becomes increasingly informative at $a = 0.8$ where their correlation strengths are almost equal. Accordingly, they then contribute comparable information to \textsc{MultiCAM}’s prediction at this epoch.

The lower panels of \autoref{fig:multicam_grid} also allow us to see how the RMSE scatter varies and compares to the variation over the range of radii. The RMSE scatter is is roughly constant over the range of $R_{\rm sp\star, 1}$. There is slightly suppressed scatter at late epochs and high $m(T)$ and also at early epochs and low $m(T)$ as a result of the non-Gaussian joint distribution of the stellar splashback features and the differential MAH at these epochs where the predictions approach the bounds of their allowed range at $m(T)=0$ and $m(T)=1$.

In total, the two stellar splashback radii have significant predictive power about the continuous MAHs of individual clusters. The \textsc{MultiCAM} framework allows us to make predictions of these MAHs with an accuracy that is conservatively characterized by the scatter in the predictions. These results indicate the potential of future measurements of the stellar splashback radii for constraining the continuous MAHs over a range of timescales for individual clusters.

\section{Discrete merger history}
\label{sec:discrete}

In addition to the established relation between smooth accretion rates and dark matter splashback features, \citet{Zhang2021} determined that major mergers also lead to a significant decrease in the dark matter splashback radius in idealized merging cluster simulations. Therefore, beyond probing the generic mass assembly of a cluster, the ICL's stellar splashback features could specifically inform discrete merger events in the cluster's past.  
In this section, we test the ability of the location of the stellar splashback radii and other cluster properties to classify clusters that have recently undergone major mergers. This analysis corresponds to the toy illustrations in right panel of \autoref{fig:flowchart}. We find that the stellar splashback radii are able to classify clusters that have or have not experienced major mergers as a function of the time that the mergers took place. We consider this classification with single properties as well as in the multivariable parameter space of the primary and secondary splashback.

Mergers are defined in the simulation as the crossing of $R_{\rm vir}$ of a host halo by another halo. 
We select major mergers to be mergers with a mass ratio to the host greater than or equal to 0.25. We measure time in units of dynamical times as defined in \autoref{eq:tdyn} and \autoref{eq:T}. 
\citet{Zhang2021} showed that the contraction of the dark matter splashback radius in response to idealized major mergers has an extended duration of approximately 0.5 to 1 dynamical times depending on the merger mass ratio. In order to probe this potential extended contraction of the stellar splashback radii, we use wide time bins with a width of 0.75 dynamical times centered on each $T$ and classify whether or not each cluster had a merger within that dynamical time window. In \autoref{app:merger_widths}, we show our results are similar for merger time windows of width 0.5, 0.75, and 1.0 dynamical times, and therefore proceed with 0.75 dynamical times for all our cluster properties. Future work with larger sample sizes of clusters could better distinguish whether these different cluster properties probe different merger time window widths.

We use the Matthews correlation coefficient (MCC) \citep{matthews1975, baldi2000assessing} to quantify how well a given cut in the properties or combination of properties classifies clusters as merged or non-merged. This metric is a discrete analog to the traditional correlation coefficient and is defined as:
\begin{equation}
\label{eq:MCC}
    \small{MCC = \frac{TP \cdot TN - FP \cdot FN}{\sqrt{(TP + FP) (TP + FN)(TN + FP)(TN + FN)}},}
\end{equation}
where $TP, TN, FP, FN$ are true positives (merged clusters classified as merged), true negatives (non-merged clusters classified as non-merged), false positives (non-merged clusters classified as merged), and false negatives (merged clusters classified as non-merged), respectively. The minimum value of the MCC, corresponding to the completely incorrect classification, is -1. The maximum value of the MCC, corresponding to perfect classification is +1. A value of zero corresponds to predictions which are completely uncorrelated from true values.
\begin{figure}[t]
\includegraphics[width=.91\columnwidth]{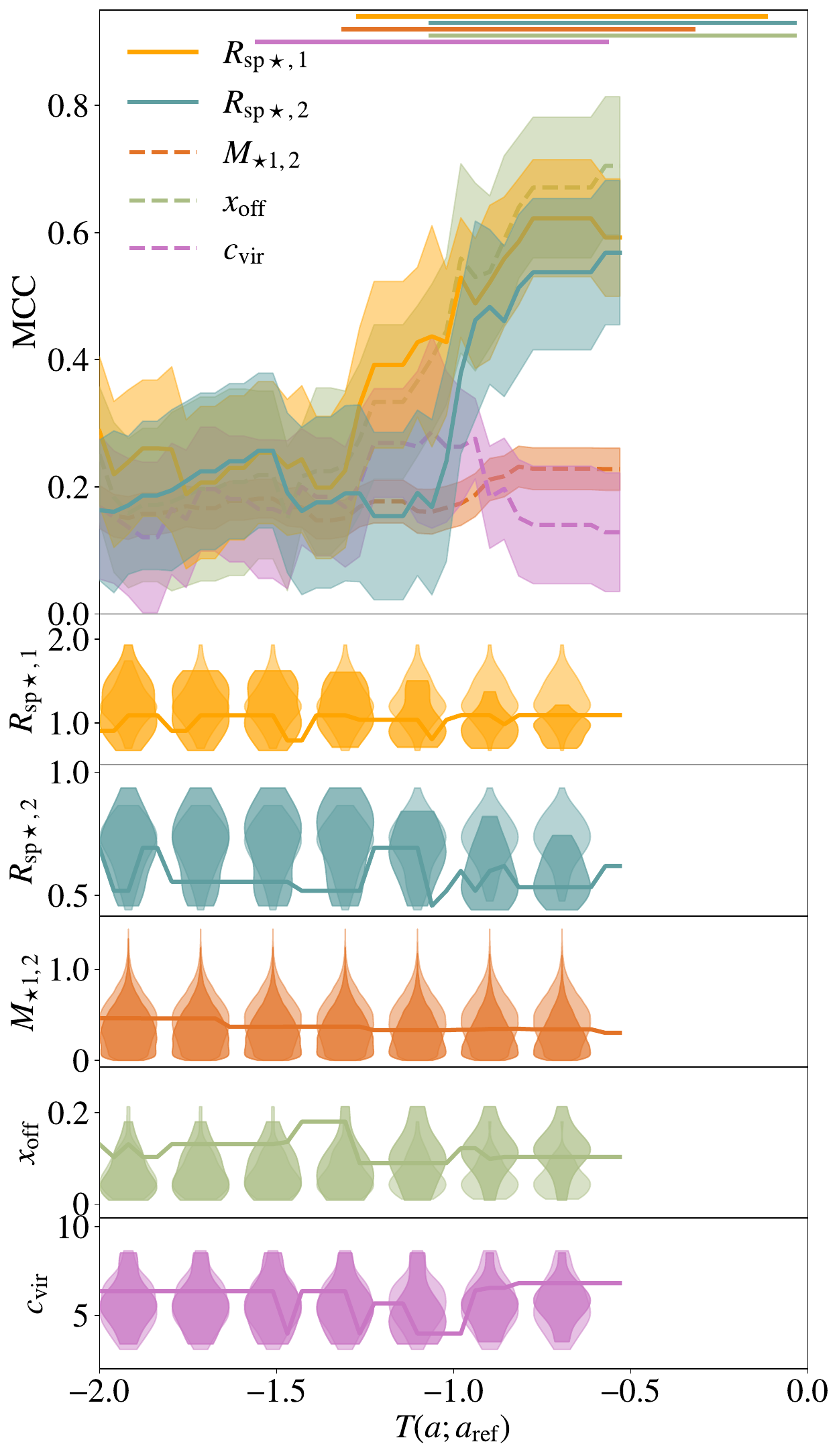}
    \caption{For each present-day cluster property, curves and violin plots illustrating the timescales at which major mergers most affect the property, thereby affording the property the most classification power about mergers at that time. Upper panel: Matthews correlation coefficients (MCC) of cuts in cluster properties optimized to best classify clusters as either having had at least one major merger or no major mergers in time windows of width 1.0 dynamical time centered on $T$. 
    The shaded bands indicate the jackknife-estimated standard error. Horizontal line segments at the top of the plot indicate the time windows corresponding to the most ``impactful`` cut, the one which maximizes the MCC between the the true and expected merger classifications.
    Lower panels: Violin plots showing the distributions of each present-day property for clusters that experienced a merger around time $T$ (dark) and for those which did not (light). The solid lines indicate the optimized cut between the distributions. When two violins are well separated, this means that the present-day property is impacted by the merger history at that time and that the optimized cut between them will correspond to a higher MCC. 
    The tracers of merger history proposed in this work, $R_{\rm sp\star, 1}$ and $R_{\rm sp \star, 2}$, are similarly capable of classifying clusters that have merged and not merged at slightly different timescales. They are slightly less capable than the known tracer of relaxedness, $x_{\rm off}$, but substantially more capable than $c_{\rm vir}$ and $M_{\star \rm 1,2}$.}
    \label{fig:classification}
\end{figure}

\subsection{Single Variable Classification of Discrete Merger History}
\label{subsec:1D}
We first test how predictive the stellar splashback radii and other tracers are of the occurrence/non-occurrence of recent major mergers. Given a distribution of present-day cluster properties for our cluster sample, for each time, $T$, we can label each property as belonging to a cluster that had at least one major merger in a window of 1 dynamical time around $T$ or belonging to a cluster that had no major mergers during that period. For each $T$, we can then determine a cut in the property that best separates clusters that did and did not merge around time $T$. This classification cut is optimized by maximizing the MCC defined in \autoref{eq:MCC}. In the uppermost panel of \autoref{fig:classification}, for each cluster property, we plot the MCC that quantifies how well an optimized cut in that property classifies the occurrence of mergers in a time window centered on a given $T$ dynamical times in the past. 
In the lower panels, for evenly sampled values of $T$, we plot the distributions of properties belonging to clusters that did or did not merge with the optimized cut indicated with a line. The optimized cut is occasionally shifted all the way to the tails of the distributions coincident with low MCC values, indicating that it failed to find a predictive cut between the distributions. In both the upper and lower panels, we do not plot the curves or violin plots for timescales more recent than 0.5 dynamical times ago. This timescale is half the width of the time window and marks the last timescale before the windows would begin reducing in width as they approach the boundary of $T = 0$.

The primary splashback radius $R_{\rm sp\star, 1}$  and the secondary splashback radius  $R_{\rm sp \star, 2}$ are both capable of separating merged clusters from non-merged clusters over a range of merger time windows with peak MCC's of $\sim0.6$. $R_{\rm sp\star, 1}$ is most sensitive to and therefore most capable of classifying clusters that underwent major mergers in the past $\sim1.3$ dynamical times. This time window is longer than the others as there is a flat range of maxima as opposed to a single peak. This timescale is consistent with the picture from idealized mergers and their impact on the dark matter primary splashback radius in \citet{Zhang2021}. This work found that mergers with a range of merger mass ratios between 0.1 to 1 lead to a strong decrease in the splashback radius with the maximal decrease offset by approximately 0.5 dynamical times from the major merger (this offset and the degree of reduction decreases with lower merger mass ratio). The reduction in the splashback radius lasts up to roughly one dynamical time from the time of their idealized major mergers. This agrees with the wide time window of mergers to which we determine $R_{\rm sp \star, 1}$ has the most sensitivity, reflecting that the present-day splashback radius is still recovering from mergers that took place around one dynamical time ago while it has not had time to contract due to the most recent major mergers. $R_{\rm sp \star, 2}$ is most capable of classifying clusters that underwent major mergers in the last dynamical time, slightly shifted from $R_{\rm sp \star, 1}$. These slightly different timescales reflect the same difference seen in \autoref{fig:correlation}. In the lower panels of \autoref{fig:classification}, It is possible to see the merged and non-merged distributions of clusters separating for $R_{\rm sp\star, 1}$ and $R_{\rm sp \star, 2}$ at the relevant timescales discussed above, where the present-day cluster property has distinguishing power.

 A tracer commonly used to characterize the dynamical state or relaxedness of a cluster, $x_{\rm off}$, is a slightly better classifier compared to the stellar splashback radii and peaks at similar timescales to $R_{\rm sp \star, 2}$. $c_{\rm vir}$ and particularly $M_{\star \rm 1,2}$ prove to be less capable at classifying merged and non-merged clusters as compared to the other tracers. The concentration, $c_{\rm vir}$, probes a slightly earlier timescale than the other tracers, mergers between 1.5 and .7 dynamical times in the past. These epochs closely match the timescales of maximal response of halo concentration to major mergers found in \cite{KWang}.  

\begin{figure}[ht]
\includegraphics[width=\columnwidth]{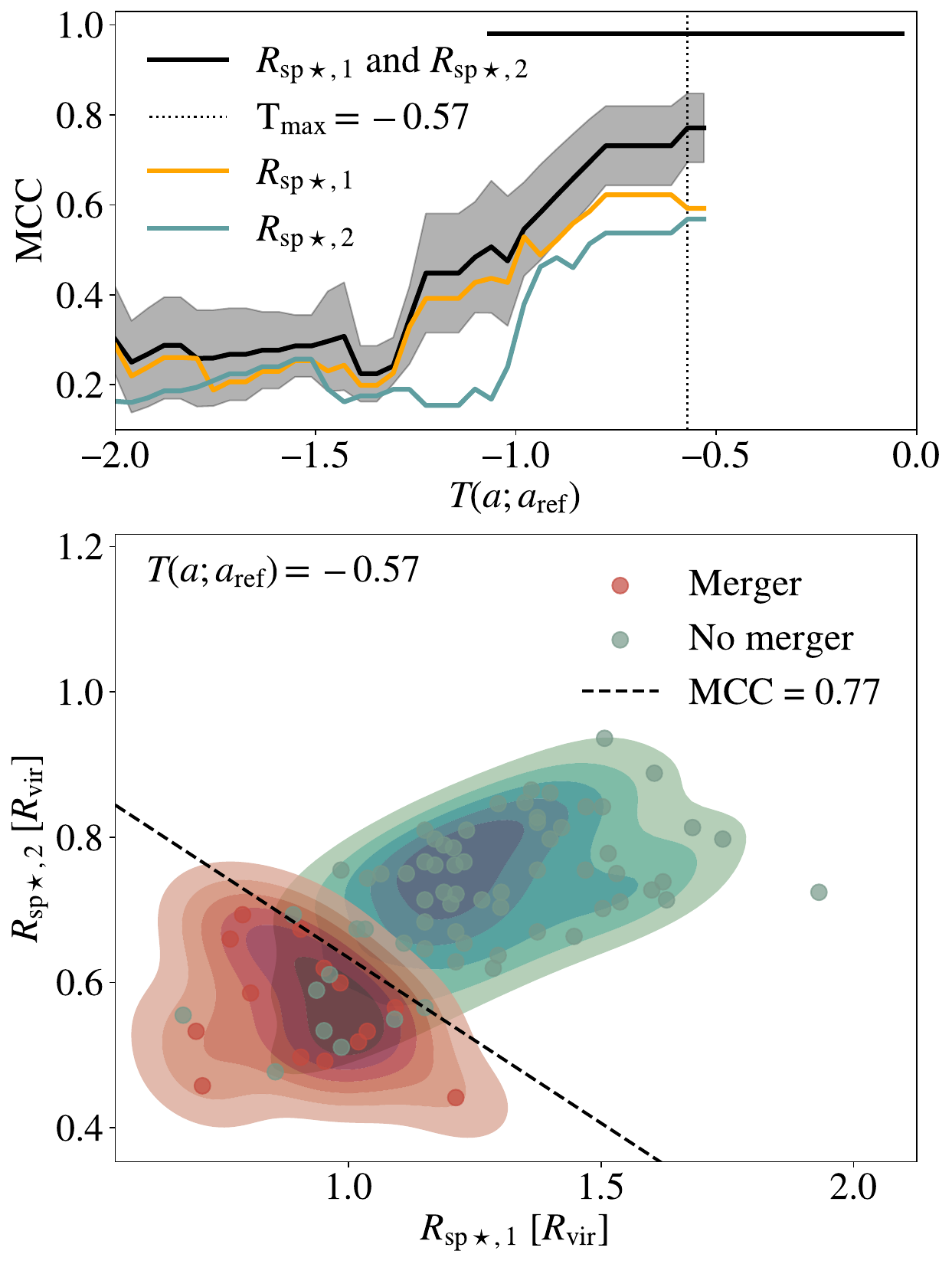}
    \caption{Upper panel: Matthews correlation coefficient (MCC) of the optimized linear cut best separating clusters that merged and did not merge in a window of 1.0 dynamical times around $T$. The shaded band indicates the standard error on the MCC as determined by jackknife resampling. The vertical dotted line indicates the timescale at which the MCC curve begins to peak ($T = -0.57$). For comparison, the MCC curves of $R_{\rm sp\star, 1}$ and $R_{\rm sp \star, 2}$ from \autoref{fig:classification} are overplotted. Lower panel: The scattered points and contours corresponding to the $R_{\rm sp\star, 1}$ and $R_{\rm sp \star, 2}$ values of clusters that merged and did not merge around that peak time of $T = -0.57$. The contours indicate six linearly-spaced levels of density of the scatter points in the space, found using a kernel density estimator. The black dashed line indicates the optimized linear cut between the red and green distributions with the highest MCC.}
    \label{fig:classification2D}
\end{figure}

\subsection{Multivariable Classification of Discrete Merger History}
\label{subsec:2D}
Analogously to combining the primary splashback radius $R_{\rm sp\star, 1}$ and the secondary splashback $R_{\rm sp \star, 2}$ radius in the \textsc{MultiCAM} framework to predict the continuous MAHs, we test how predictive the stellar splashback radii are in combination for the occurrence of recent major mergers. At a range of epochs, we determine the optimized linear cut between the clusters that merged and did not merge in the multivariable space of $R_{\rm sp\star, 1}$ and $R_{\rm sp \star, 2}$. The upper panel of \autoref{fig:classification2D} is analogous to the upper panel of \autoref{fig:classification}. As a function of dynamical times $T$, we show the optimized MCC of the line separating clusters that merged and did not merge in a window of 1.0 dynamical times around $T$. We demarcate the peak of the MCC at $T=-0.57$ as a vertical dotted line. The time window corresponding to this peak is indicated with a horizontal black line. We overplot the MCC curve of $R_{\rm sp\star, 1}$ and $R_{\rm sp\star, 2}$ alone from \autoref{fig:classification} for comparison. In the lower panel, we show the distributions of merging and non-merging clusters in this multivariable space of $R_{\rm sp\star, 1}$ and $R_{\rm sp \star, 2}$ around that peak epoch of $T=-0.57$.  

At the peak epochs, corresponding to sensitivity to mergers in the last $\sim 1$ dynamical time, the combination of $R_{\rm sp\star, 1}$ and $R_{\rm sp \star, 2}$ substantially improves the classification of merging and non-merging cluster populations over each stellar splashback radii alone. Importantly, the combination of the stellar splashback radii achieves better classification capability than the traditional probe of relaxedness, $x_{\rm off}$. Thus, if observable, the stellar splashback radii offer an independent avenue for determining the dynamical state of clusters using optical data.

\section{Observability of ICL at Stellar Splashback Radii}
\label{sec:obs}

We now investigate the observability of the primary and secondary stellar splashback radii with future optical and infrared (IR) surveys. In order to determine the radial extent of the ICL potentially measurable given the predicted surface brightness limits of specific surveys, we estimate a 2D projected surface brightness profile for each simulated cluster given its 3D stellar mass distributions and estimated mass-to-light ratios. For a given survey band, we determine the mass-to-light-versus-color relation (MLCR) in the relevant band and combine with observed ICL colors to obtain a fiducial mass-to-light ratio along with upper and lower mass-to-light ratios from the observed error bars on the colors. We detail these methods and the specific data and models used in \autoref{app:SB}. 

\begin{figure}[t]
\includegraphics[width=\columnwidth]{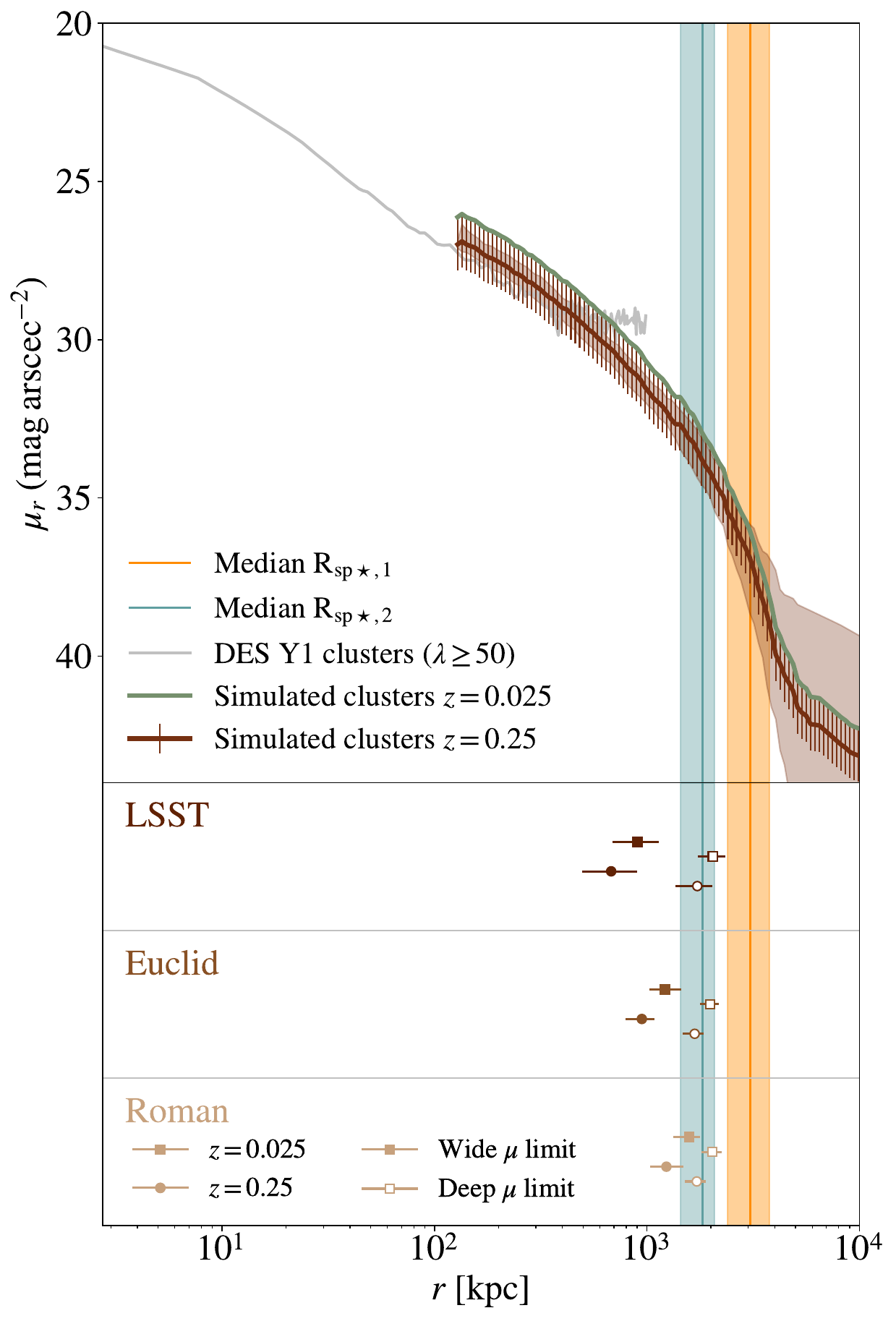}
    \caption{Upper panel: Median $r$-band projected surface brightness profiles of our simulated clusters ``observed'' at $z = 0.25$ along with a high richness sample of DES Y1 clusters at $z = 0.25$ from \citet{Zhang2019_DES}. The solid brown curve shows the median simulated cluster profiles using the fiducial $M$/$L_{r}$ with error bars showing the range of surface brightnesses allowed by the error bars on the observed ICL colors used to obtain the $M$/$L_{r}$. The shaded band around the solid curve spans the $16^{\rm th}$ to $84^{\rm th}$ percentiles of the simulated cluster sample. We further show a lower redshift $z = 0.025$ median surface brightness profile using the fiducial $M$/$L_{r}$. The median and $16^{\rm th}$ to $84^{\rm th}$ percentiles of $R_{\rm sp\star, 1}$ and $R_{\rm sp \star, 2}$ for the cluster sample are indicated as vertical lines and shaded bands. Lower panel: Radial limits of the indicated surveys calculated by comparing the surface brightness limits for each survey to the median surface brightness profile in the relevant band. Each point indicates the radial limit using the fiducial $M$/$L$ with error bars propagating the error on the $M$/$L$. Radial limits are shown corresponding to both wide and deep surface brightness limits as well as for $z = 0.025$ and $z = 0.25$. Lower redshift clusters allow for a small gain in probing the outskirts of the ICL. These limits in radial reach of these surveys indicate that $R_{\rm sp \star, 2}$ is within the reach of these surveys, particularly the Nancy Grace Roman Space Telescope, while $R_{\rm sp \star, 1}$ is potentially too faint to observe with the existing survey strategies.}
    \label{fig:SB}
\end{figure}

In the upper panel of \autoref{fig:SB}, we show the median surface brightness profile in the $r$-band at a redshift of $z = 0.25$ for our sample of clusters. The solid curve indicates the profile made using our fiducial $r$-band mass-to-light ratio ($M$/$L_{r} = 2.39$) with error bars showing the range of surface brightnesses allowed by the error bars on the observed ICL colors corresponding to $M$/$L_{r} = 1.11$ and $M$/$L_{r} = 5.13$ (see \autoref{app:SB} for further details). We additionally show the median fiducial profile at a redshift of $z = 0.025$, which marginally increases the surface brightness at all radii due to less redshift dimming.
At low radii, these curves cut off at the physical radius in kpc corresponding to $0.05R_{\rm vir}$. Within this radius, the clusters would be dominated by their BCG, which is not modeled in this work. 
We include the stacked $r$-band surface brightness profiles of the $z = 0.25$ highest richness sample of galaxy clusters from the DES Y1 release as presented in \citet{Zhang2019_DES}. This observed profile includes the effects of the BCG and due to the surface brightness limits of the DES Y1 survey, does not extend to the outer regions of the ICL.
We further display vertical lines and shaded regions that delineate the medians and $16^{\rm th}$ and $84^{\rm th}$ percentiles of $R_{\rm sp\star, 1}$ and $R_{\rm sp \star, 2}$. We note that taking the population median of the stellar density profile washes out the stellar splashback features of individual clusters, since they occur at different locations for each cluster. 

We find that our surface brightness profiles of simulated clusters are consistent with the higher richness (and more massive) sample of observed DES clusters. 
Our surface brightness profiles are comparable to the simulated projected surface brightness profiles generated by \citet{Deason_StellarSplashback} who similarly found consistency with the DES Y1 results. \citet{Deason_StellarSplashback} use a constant mass-to-light ratio of $5 M_{\odot}$/$L_{\odot}$ as opposed to our fiducial value of 2.39. While we match their surface brightness values between $\sim$300 kpc and 1 Mpc, we predict surfaces brightnesses which are $\sim$1.5 mag arcsec$^{-2}$ lower at $r\approx3$ Mpc.

\subsection{Capabilities of future surveys}
\label{sec:futureobs}

We consider the potential capabilities of three future surveys: the ground-based Vera C. Rubin Observatory's Legacy Survey of Space and Time (LSST), the Nancy Grace Roman Space Telescope High-Latitude Wide Area Survey (HLWAS), and the Euclid space telescope's Euclid Wide Survey (EWS) and Euclid Deep Survey (EDS). 
For each survey, we determine the shallow surface brightness limits that will cover the most area as well as deeper limits that will apply to limited deep fields.

LSST has been reported in the literature to have a range of values for its 10-year limiting surface brightness in the $r$-band, including $\sim$30.3 mag arcsec$^{-2}$\footnote{Note that surface brightness limits depend on the area and significance threshold considered and we report all values in this section assuming a significance criterion of 3$\sigma$ and an area of 10"$\times$10". In the case of limits reported with the criterion of 5$\sigma$, we convert to 3$\sigma$ by adopting the formula provided in \citet{Roman2020} resulting in a simple addition of 0.55 to the surface brightness limit in mag arcsec$^{-2}$. In addition to angular size and significance criteria, surface brightness limits also vary with the method of sky background subtraction \citep{Roman2020, Brough2020, Martin2022}.} \citep[e.g.][]{Brough2023}, $\sim$30.6 mag arcsec$^{-2}$ \citep[e.g.][]{Laine2018} and $\sim$32.5 mag arcsec$^{-2}$ \citep[e.g.][]{Brough2020}; $\sim$33.5 mag arcsec$^{-2}$ in the deep drilling fields \citep[e.g.][]{Brough2020}; and $\sim$34.5 mag arcsec$^{-2}$ with best-practices characterization of the sky background \citep[e.g.][]{Ji2018,Martin2022}. Given the variability in predicted limits, past works have chosen to represent the LSST surface brightness limit predictions as a wide range of 28 to 31 mag arcsec$^{-2}$ \citep{Martin2022} and 30.5 to 32 mag arcsec$^{-2}$ \citep{Deason2022_dwarfs}. For our analysis of the radial extents of massive clusters probed by LSST, we consider a ``wide" surface brightness limit in the $r$-band of 30.3 mag arcsec$^{-2}$ and a ``deep" limit of 33.5 mag arcsec$^{-2}$.
In the lower panel of \autoref{fig:SB}, we show the radial limits of LSST corresponding to where the wide and deep surface brightness limits cross the median surface brightness profiles of the simulated clusters assuming the fiducial $M$/$L_r$ value along with its error bars and two possible redshifts of $z = 0.025$ and $z = 0.25$. While the shallower wide survey is unlikely to observe out to either stellar splashback radius, the deep fields have the potential to observe $R_{\rm sp\star, 2}$. \citet{Deason_StellarSplashback} found that in order to observe $R_{\rm sp\star, 1}$, the observational data would have to go out to 32--36 mag arcsec$^{-2}$, which they claimed would be possible with LSST. We find that a range of 35--38 mag arcsec$^{-2}$ might be necessary to observe $R_{\rm sp\star, 1}$, which is likely unattainable with LSST. This difference in predictions between \citet{Deason_StellarSplashback} and this work might be due to several differences in our cluster mass samples, simulations, and splashback measurement techniques. We note that due to our definition of the splashback radius as a slope cutoff in the radial stellar mass density profile instead of a slope minima, it is possible that we overestimate the physical radius corresponding to the splashback feature and therefore slightly underestimate the brightness of the stellar splashback locations, particularly for $R_{\rm sp\star, 1}$. 

The Euclid space mission is predicted to reach $\sim$29.5 mag arcsec$^{-2}$ in the Euclid Wide Survey and $\sim$31.5 mag arcsec$^{-2}$ in the Euclid Deep Survey \citep{EuclidCollab2022}. We use these two values as our ``wide" and ``deep" surface brightness limits for Euclid in the $H_E$ band. The lower panels of \autoref{fig:SB} show that the radial limits of Euclid Deep Survey calculated for a fiducial $M$/$L_{H_E}$ value along with its error bars (see \autoref{app:SB}) and two possible redshifts allow for the observation of $R_{\rm sp\star, 2}$ although $R_{\rm sp\star, 1}$ is seemingly out of reach. 
Euclid early release observations have already reported measured surface brightness profiles of two individual clusters out to $\sim600$ kpc \citep{Kluge2024, Ellien2025}.
\citet{Bellhouse2025_Euclid} forecasted the ICL detection capabilities of the Euclid Wide Survey and predicted that the primary stellar splashback feature could be observable for individual clusters with halo mass $> 10^{14.7} M_{\odot}$ at $z < 0.3$ or by stacking clusters in bins of richness or lensing mass. Their more optimistic estimate of the observability of $R_{\rm sp\star, 1}$ is dependent on its location relative to $R_{200\rm m}$ taken from \citet{Deason_StellarSplashback}.
As mentioned above, we are more pessimistic about the observability of the stellar splashback features than \citet{Deason_StellarSplashback}.

The Nancy Grace Roman Space Telescope's HLWAS is expected to observe in the F158 or $H$-band down to surface brightness limits of $\sim$30.5 mag arcsec$^{-2}$ for the medium tier of the High Latitude Wide Area Survey (HLWAS) \citep{RomanWhitePaper, RomanTACreport2025}. The deep tier of the HLWAS has a smaller area of 19.2 deg$^2$ but reaches a fainter limit of $\sim$31.7 mag arcsec$^{-2}$ \citep{RomanTACreport2025}. We consider these two values as our ``wide" and ``deep" surface brightness limits for Roman in the $H$ band and calculate fiducial surface brightness profiles with error bars using the fiducial $M$/$L_{H}$ and its error bars that we estimate in \autoref{app:SB}. The lower panels of \autoref{fig:SB} show that the radial limits of Roman's HLWAS allow for the observation of $R_{\rm sp\star, 2}$ even in the wide tier at low redshift although $R_{\rm sp\star, 1}$ is potentially out of reach without additional data processing strategies optimized for the faint ICL. Out of the three surveys considered in this initial analysis, the Roman Space Telescope appears best situated to observe these outer regions of the ICL and potentially detect stellar splashback features.

We briefly note that an additional space-based instrument, the James Webb Space Telescope (JWST) observing in the near-infrared has been used to measure the ICL of a $z  = 0.39$ cluster out to a radial distance of $\sim$400 kpc \citep{Montes_Trujillo_2022}. JWST has also been shown to be capable of imaging diffuse light and resolved stars with NIRCam imaging and photometry down to an extremely low surface brightness in the $V$-band of $\sim$35 mag arcsec$^{-2}$ for a low-mass galaxy at a distance of 35 Mpc \citep{Conroy2024}. Thus, there is tantalizing potential for targeted JWST observations to be able to image selected regions of the outer ICL for nearby clusters such as the Virgo cluster. 

Despite global estimates of limiting surface brightness for these large surveys, careful and dedicated data processing techniques have historically been able to push these limits. For example, the IAC Stripe82 Legacy Project observed ultra-diffuse galaxies down to 28.6 mag arcsec$^{-2}$ in SDSS $r$-band using careful stacking and sky background measurements when the general limiting surface brightness was measured to be 26.2 mag arcsec$^{-2}$ \citep{Kniazev2004, Roman2017}. \citet{Chen2022} even used stacks of 3000 SDSS clusters imaged in the $r$-band to measure their diffuse light down to $\sim$32 mag arcsec$^{-2}$, corresponding to a radial distance of $\sim$1 Mpc from the center of the clusters. \citet{Montes2021} unveiled the low surface brightness ICL of clusters in archival HSC data down to $\sim30.9$ mag arcsec$^{-2}$ in $g$-band surface brightness with such techniques. Similarly, careful background estimation and systematic bias mitigation allowed \citet{Gonzalez_George_Connor_Deason_Donahue_Montes_Zabludoff_Zaritsky_2021} to observe the ICL surface brightness profile of an HST Frontier Fields cluster out to $\sim 2$ Mpc and down to $\sim32$ mag arcsec$^{-2}$ in their F105W filter, enabling a possible detection of the stellar splashback radius. A study exploring the ability of different data sets to study stellar halos of massive galaxies stressed the importance of the treatment of the sky background and whether or not this treatment and empirical sky corrections occur on the level of coadd images or within the data reduction pipelines. In the context of large surveys, care must be taken to avoid global data reduction and background subtraction pipelines optimized for point source detection rather than low surface brightness structures \citep{Li2022}. The flexibility to work with images before global sky subtraction and coadding, along with customized data reduction pipelines for low surface brightness science could allow studies of the ICL with LSST and other surveys to push the boundaries of what is considered observable.  

In recent years, targeted ground-based surveys have revolutionized our capabilities of observing faint systems such as galactic stellar halos, tidal features, and dwarf galaxies. The Dragonfly Telephoto Array and its resulting surveys including the Dragonfly Edge-on Galaxy Survey, the Dragonfly Wide Field Survey, and the Dragonfly Ultrawide Survey, employ a novel telescope design and dedicated data reduction pipelines optimized for the study of extended low surface brightness structures \citep{Gilhuly2022_Dragonfly, Shen2024_Dragonfly}. The LIGHTS survey performed ultra-deep imaging with the Large Binocular Telescope (LBT) along with attentive observing data reduction procedures that addressed variable observing conditions, scattered light, and other observational challenges to achieve limiting surface brightnesses deeper than $\sim$30.5 mag arcsec$^{-2}$ in the $r$-band in their study of stellar halos and faint satellite galaxies \citep{Trujillo2021_LIGHTS,Zaritsky2024_LIGHTS}. The Korea Microlensing Telescope Network (KMTNet) was used to conduct a deep and wide-field survey of nearby galaxies with careful sky subtraction to preserve low surface brightness features \citep{Byun2022_KMTNet}. These and many other targeted surveys demonstrate the power of employing technology and data processing methods specifically optimized for the study of the low surface brightness universe. Similar approaches can be considered for the wide field universe at the scale of intracluster light in the future. 

In total, our initial exploration into the observability of the stellar splashback features suggest that outer ICL radii occupied by $R_{\rm sp\star,2}$ will be observable with forthcoming surveys from LSST, Euclid, and, particularly, Roman while $R_{\rm sp\star,1}$ likely lies below their nominal surface brightness limits. However, observational strategies, data reduction pipelines, and even future telescope designs tailored for low surface brightness studies, such as those used in Dragonfly or LIGHTS, may considerably improve the observational prospects of these faint ICL features. 

\subsection{Considerations for detectability of stellar splashback features}
\label{subsec:detectability}

We have examined whether the radial regions of the ICL that we determine to host the 3D radii stellar splashback features will be observable with various surveys and strategies. However, we emphasize that the observability of the ICL surface brightness profile at these radii does not necessitate the detectability of the stellar splashback features. Our analysis has relied on idealized methods of obtaining these stellar splashback radii using a combination of particle orbits and angular median radial stellar mass density profiles. Several effects can impact the detectability of the stellar splashback features using minima in the logarithmic derivatives of radial surface brightness profiles. The recent mass accretion history can impact the depth of the secondary stellar splashback logarithmic derivative minima in addition to their radial location \citep{Deason_StellarSplashback, Adhikari2014}. In order to determine the detectability of the secondary stellar splashback radii as a function of mass accretion rate, it is further necessary to include observational effects and develop analyses that can differentiate between real stellar splashback minima and noise in the observed projected surface profiles. Though this work focuses on an initial idealized exploration of the information content and observability of these ICL outskirts, future work can forecast the detectability of these stellar splashback features and the interplay with mass accretion rates, profile analysis methods, observational uncertainties, and future survey properties.

\section{Conclusions}
\label{sec:conclusions}

The mass accretion histories of massive galaxy clusters are chronicled in their intracluster light (ICL). In this study, we explore how much of this history can be gleaned from the stellar splashback features of these clusters. 
We leverage a combination of the high-resolution dark matter-only Symphony simulation suite and the \textsc{Nimbus} star-tagging model to make detailed predictions for the structure of the ICL for a sample of 79 massive clusters. For each cluster, we make idealized measurements of the present-day primary and secondary stellar splashback radii ($R_{\rm sp\star, 1}$ and $R_{\rm sp \star, 2}$), corresponding to the edges of the radial distribution of stars that have passed their first and second orbital pericenters respectively. The dark matter counterpart to $R_{\rm sp\star, 1}$ is known to strongly relate to the accretion rate of a cluster. The dark matter counterpart to $R_{\rm sp \star, 2}$, often known as the ``second caustic,'' has been understood to be distinguishable at low accretion rates but is often washed out by the stacking of dark matter profiles \citep[e.g.,][]{Xhakaj_Diemer_Leauthaud_Wasserman_Huang_Luo_Adhikari_Singh_2020, Deason_StellarSplashback}. We further measure and compare to additional known tracers of the formation history and dynamical state of clusters.

We first examine how individually informative the present-day stellar splashback radii are about the continuous mass accretion histories (MAH) of individual clusters:
\begin{enumerate}[i]
\item $R_{\rm sp\star, 1}$ and $R_{\rm sp \star, 2}$ are both highly correlated with the mass of the cluster at a range of past epochs. $R_{\rm sp\star, 1}$ contains the most information about the mass one dynamical time ago while $R_{\rm sp \star, 2}$ probes a more recent epoch, $\sim0.7$ dynamical times ago (see \autoref{fig:correlation}).  
\item The correlation strength and peak timescale of $R_{\rm sp\star, 1}$ and $R_{\rm sp \star, 2}$ are comparable to the known formation history tracer $x_{\rm off}$, the offset between the halo center of mass and its most bound particle, but are considerably more correlated with the continuous MAH than the concentration $c_{\rm vir}$ and the stellar mass gap $M_{\star 1,2}$.
\end{enumerate}

We subsequently examine how well the combined stellar splashback information allows for the reconstruction of the full continuous MAH:
\begin{enumerate}[i]
\item Using the \textsc{MultiCAM} framework, we construct an explicit model that can generate mock MAHs given the stellar splashback radii. 
\item The radii $R_{\rm sp\star, 1}$ and $R_{\rm sp \star, 2}$ probe different epochs of the mass accretion history, and are, in combination, capable of generating realistic mock MAHs (\autoref{fig:MAH} and \autoref{fig:multicam_grid}).
\end{enumerate}

We then investigate how predictive the stellar splashback radii are about the discrete merger history of a cluster by measuring how well each feature individually classifies the occurrence of recent major mergers:
\begin{enumerate}[i]
\item $R_{\rm sp\star, 1}$ and $R_{\rm sp \star, 2}$ are both able to classify the occurrence of recent mergers within the last $\sim 1$ dynamical time with a classification metric slightly lower than $x_{\rm off}$ but considerably better than $c_{\rm vir}$ and $M_{\star 1,2}$ (\autoref{fig:classification}). 
\item A linear cut in the multivariable space of $R_{\rm sp\star, 1}$ and $R_{\rm sp \star, 2}$ substantially improves the classification, making the combination a better classifier than $x_{\rm off}$ (\autoref{fig:classification2D}).
\end{enumerate}

Lastly, we situate these results in the context of what outer regions of the cluster might or might not be observable in the next generation of optical and IR surveys:
\begin{enumerate}[i]
\item Our results suggest $R_{\rm sp \star, 2}$ is potentially observable by LSST, Euclid, and, in particular, Roman. 
\item $R_{\rm sp\star, 1}$ might be beyond these survey's capabilities given current estimates of limiting surface brightnesses (\autoref{fig:SB}).
\end{enumerate}

In this work, we have shown that the stellar splashback features of the intracluster light (ICL) encode rich information about the growth histories of galaxy clusters. Using high-resolution Symphony simulations combined with the \textsc{Nimbus} star-tagging model, we find that the primary ($R_{\rm sp\star,1}$) and secondary ($R_{\rm sp\star,2}$) stellar splashback radii trace both continuous mass accretion and discrete merger events over the past few dynamical times. These features outperform traditional indicators such as the stellar mass gap and halo concentration, and are comparable to commonly used X-ray-based tracers of dynamical state. Leveraging the \textsc{MultiCAM} framework, we demonstrate that these measurements can be used to reconstruct realistic cluster mass accretion histories across a wide range of timescales.

While our analysis is idealized, these results suggest that stellar splashback features --- especially at large radii --- could serve as visible tracers of structure formation, potentially accessible to next-generation surveys such as Rubin, Roman, and Euclid. With sufficient imaging depth and control of systematics, the diffuse starlight of the ICL may become a powerful probe of the unseen, underlying dark matter distribution and the signatures imprinted within it. Though it is inevitably an incomplete record of a cluster’s life and the events that impact it, the ICL and its luminous density features express the personal story of each cluster---a memoir of mass accretion.

\section{Acknowledgements}
We thank Sidney Mau for helpful discussions about ICL observability, Benedikt Diemer and Susmita Adhikari for comments on a draft, and the GFC@Stanford group for many helpful discussions.
This work used data from the Symphony suite of simulations, hosted at \url{http://web.stanford.edu/group/gfc/gfcsims}; we thank our Symphony collaborators for their contributions to this work and making the data available. This work was supported by the Kavli Institute for Particle Astrophysics and Cosmology at Stanford University and SLAC National Accelerator Laboratory, and by the U.S. Department of Energy under contract number DE-AC02-76SF00515 to SLAC National Accelerator Laboratory.

This work received support from the U.S. Department of Energy under contract number DE-AC02-76SF00515 to SLAC National Accelerator Laboratory, and from Stanford University. This research made use of computational resources at SLAC National Accelerator Laboratory, a U.S.\ Department of Energy Office, and at the Sherlock cluster at the Stanford Research Computing Center (SRCC); the authors are thankful for the support of the SLAC and SRCC computational teams.

\appendix

\renewcommand*\thefigure{\thesection\arabic{figure}}
\setcounter{figure}{0}

\section{Sensitivity of the stellar splashback radii to the size--virial radius relation}
\label{app:Rhalf}
In this section, we test the sensitivity of our measured stellar splashback radii to the stellar size--virial radius relation used in our galaxy--halo connection model. This is important to check because we use a fixed slope cutoff of the radial stellar mass density profiles to determine the stellar splashback radii. So it is possible, in principle, that $R_{\rm sp\star, 1}$ and $R_{\rm sp \star, 2}$ could correlate with galaxy sizes. We find that the stellar splashback radii are only mildly sensitive to even extreme changes in $r_{\star}$/$r_{\rm vir}$.

\begin{figure}[t]
\includegraphics[width=\columnwidth]{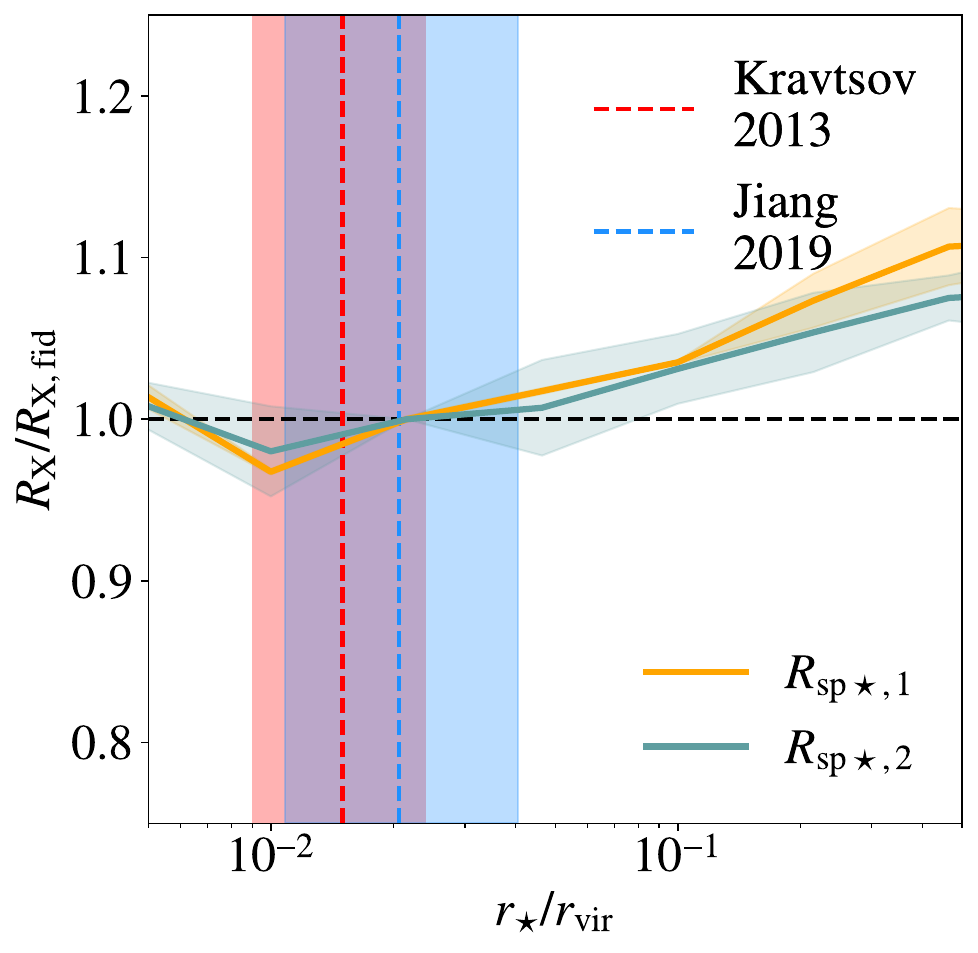}
    \caption{For an extreme range of fixed size--virial radius relations, the median ratio of the stellar splasback radii, $R_{\rm sp\star, 1}$ and $R_{\rm sp \star, 2}$, to their fiducial values using the \citet{Jiang2019} relation. The shaded regions around the curves indicate the jackknife standard error in those medians. The vertical blue dashed line and shaded band indicate the median and 68\% scatter of the $r_{\star}$/$r_{\rm vir}$ assigned to our galaxies using \citet{Jiang2019} relation. The vertical red dashed line and shaded band indicate the preferred $r_{\star}$/$r_{\rm vir}$ determined by \citet{Kravtsov2013} along with the reported scatter. Both $R_{\rm sp\star, 1}$ and $R_{\rm sp \star, 2}$ show only a slight trend with increasing $r_{\star}$/$r_{\rm vir}$ and are consistent with the fiducial values in the regions spanned by the \citet{Kravtsov2013} and \citet{Jiang2019} relations.}
    \label{fig:Rhalf}
\end{figure}

In \autoref{fig:Rhalf}, we display how our measured $R_{\rm sp\star, 1}$ and $R_{\rm sp \star, 2}$
vary over a wide range of fixed size--virial radius relations as compared to their fiducial values using \autoref{eq:Jiang}. 
Both $R_{\rm sp\star, 1}$ and $R_{\rm sp \star, 2}$ are only lightly sensitive to the extreme changes in the size--virial radius relation with effects under 10\%. This trend is particularly negligible in the blue and red shaded regions occupied by the $r_{\star}$/$r_{\rm vir}$ values assigned to our galaxies with the fiducial \citet{Jiang2019} relation (see \autoref{eq:Jiang}) and the \citet{Kravtsov2013} relation, respectively.

\setcounter{figure}{0}
\section{Sensitivity of results to stellar splashback definitions}
\label{app:spdef_tests}

\begin{figure}[t]
\includegraphics[width=\columnwidth]{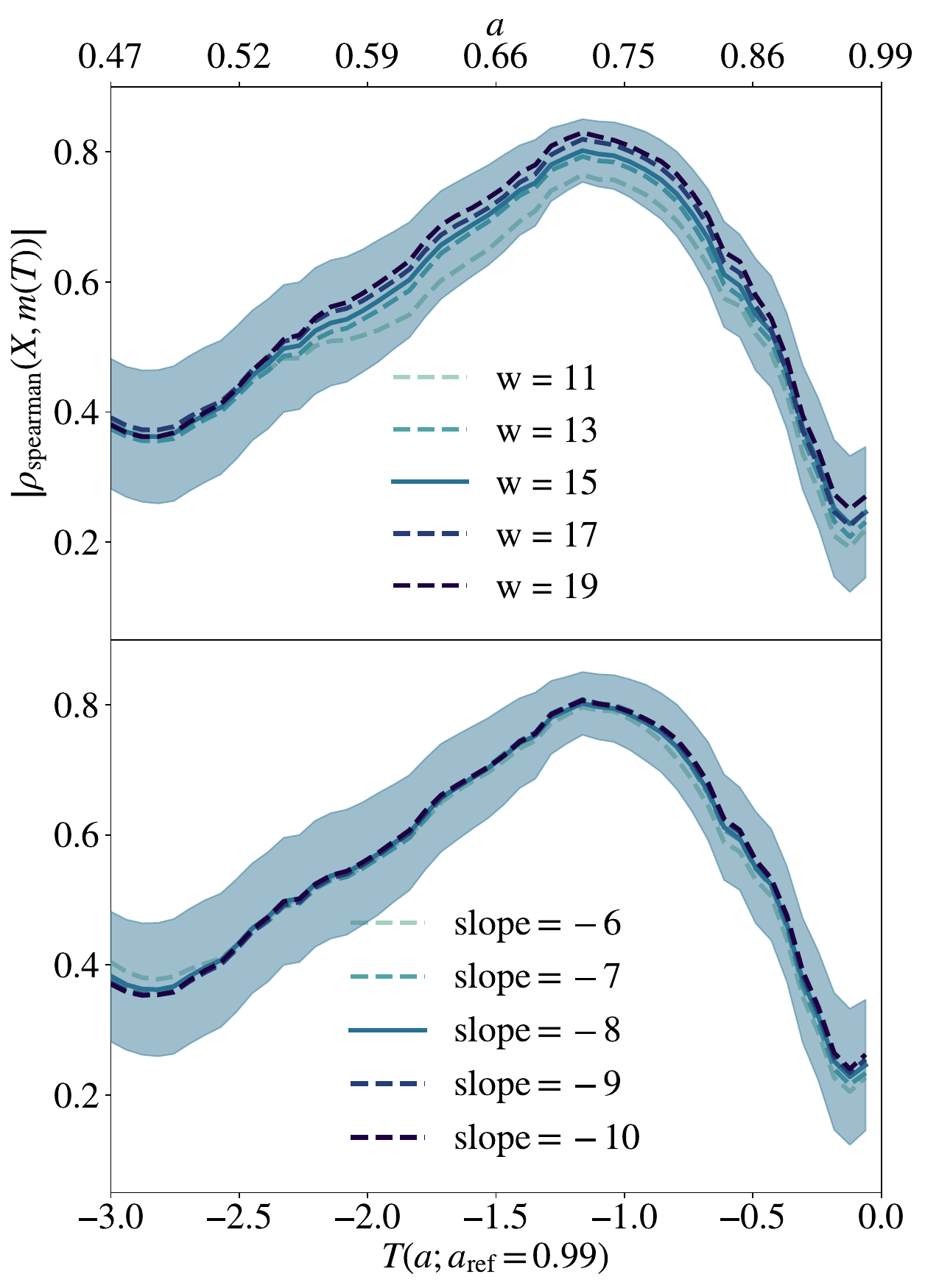}
    \caption{Analogous to \autoref{fig:correlation}, the Spearman's correlation coefficient between $R_{\rm sp \star,1}$ and the MAH varying two choices in our stellar splashback radii measurements. Upper panel: Varying the window size of the Savitsky-Golay smoothing filter between 11 and 19 radial bins. Lower panel: Varying the slope cutoff in the radial stellar mass density profile used to define the location of the stellar splashback radii. Our fiducial choices are displayed as solid lines while others are dashed. We show the jackknife standard error as a shaded band around the fiducial curve. We show a slight trend with window size that remains within the standard error shaded band. We are entirely insensitive to changes in the slope cutoff.}
    \label{fig:tests}
\end{figure}

In this section, we test the sensitivity of our results to the definitional choices used in our measurements of the stellar splashback radii as described in \autoref{subsec:MeasureRsp}. These choices are namely the Savitzky-Golay window size used in calculating the log derivatives of the radial stellar mass density profiles and the cutoff in the log derivative at which we define the stellar splashback radii. In \autoref{fig:tests}, we vary these choices in our measurement of $R_{\rm sp \star,1}$ for each cluster and present an analogous figure to \autoref{fig:correlation}. 

In the upper panel of \autoref{fig:tests}, we vary the window size of the Savitsky-Golay smoothing filter between 11 and 19 radial bins corresponding to a range of $\sim0.37$ to $\sim0.65$ dex in $r$/$R_{\rm vir}$. Our fiducial choice is 15 bins, which we indicate as a solid line with a shaded band showing the jackknife standard error. We find that the correlation strengths of $R_{\rm sp \star,1}$ with the continuous MAH show a slight trend with the window size with our fiducial choice of 15 bins falling in the middle. The spread of this trend remains within the standard error shaded band of that fiducial choice. We emphasize that following this test, we continue using the window size of 15 bins and do not fine-tune our method to increase the correlations of $R_{\rm sp \star,1}$.

In the lower panel of \autoref{fig:tests}, we vary the radial stellar mass density profile slope cutoff used to define the location of the stellar splashback radii, where the slope is the log derivative of the profile. Our fiducial choice is $-8$ and we vary between $-6$ and $-10$. The correlation curve is entirely insensitive to this variation in slope cutoff.

\setcounter{figure}{0}
\section{\textsc{MultiCAM} scatter as a measure of accuracy}
\label{app:scatter}

\begin{figure}[h]
\centering
\includegraphics[width=\columnwidth]{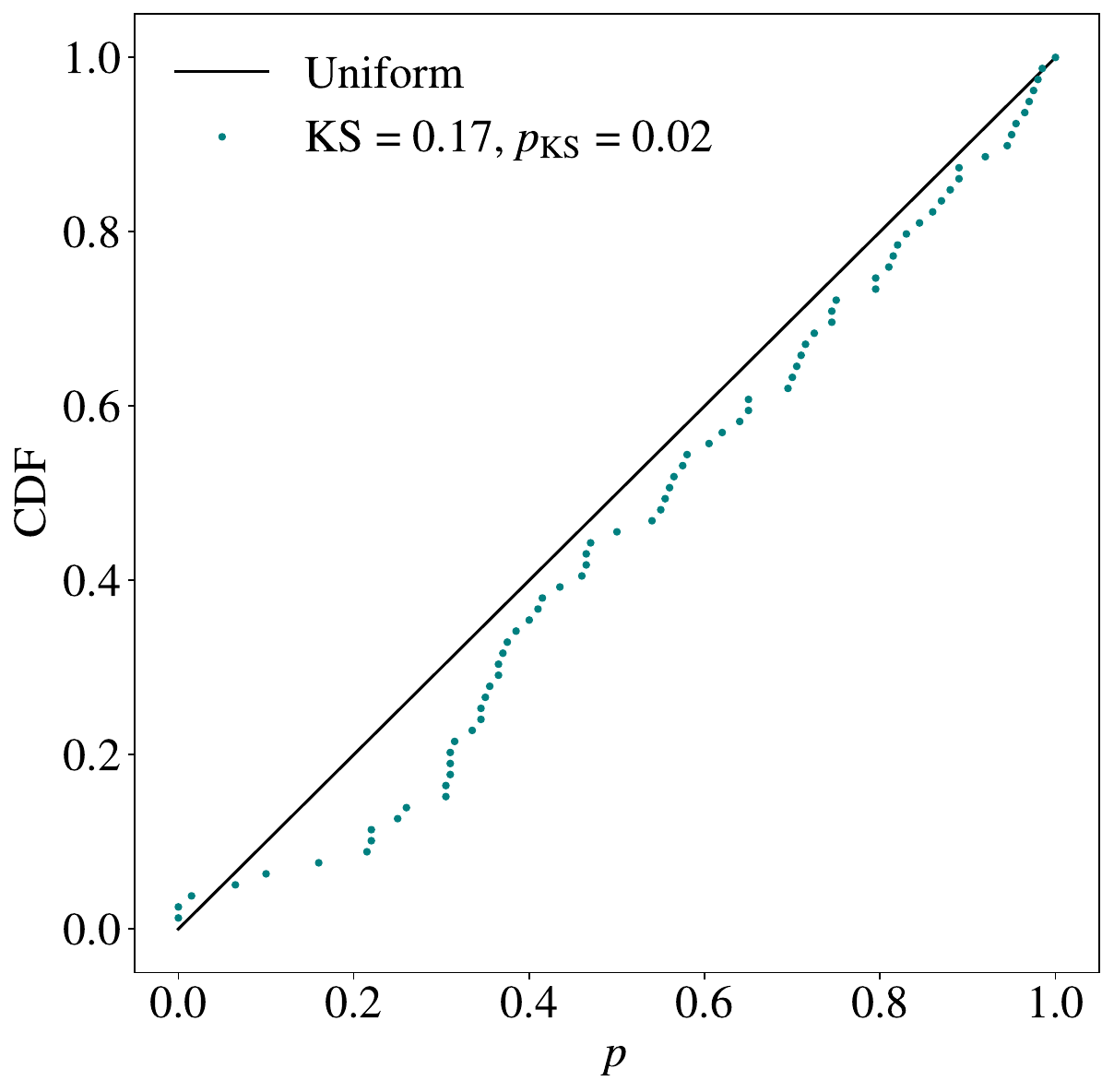}
    \caption{Cumulative distribution function (CDF) of clusters' $p$ values where $p$ value is defined as the fraction of predicted sample MAHs that have a higher RMSE when compared to the median predicted sample MAH than the true MAH (see the text for further details). The solid black line indicates the CDF of a uniform distribution for comparison. As the CDF lays below the uniform distribution, \textsc{MultiCAM}'s scatter in predicted sample MAHs is a conservative description of the accuracy of its median predicted MAH.}
    \label{fig:p}
\end{figure}

In this section, we discuss the evidence for the claim that \textsc{MultiCAM}'s predictions of the MAH are conservatively characterized by the scatter in those sample predictions. By ``conservatively characterized,'' we mean to test whether the scatter in MultiCAM predictions for fixed input parameters is an upper bound on the distribution of expected deviations between the median relation and the truth for fixed input parameters. We note that these two quantities will be the same by construction if \textsc{MultiCAM}'s ansatz that the predictor and target variables are related by a multi-dimensional Gaussian copula (i.e. a distribution which can be transformed into a Gaussian by remapping the marginal distributions) is correct.

For each cluster, we calculate the root mean squared error (RMSE) of each sample prediction with respect to the median of 200 sample predictions for a scale factor range of $a = 0.6$ to $a = 0.9$. We compare this RMSE of the samples to the RMSE between the true MAH and the median of the samples. We define a $p$ value to be the fraction of sample MAHs that have a higher RMSE than the true MAH. In \autoref{fig:p}, we show the cumulative distribution function of $p$ values for all the clusters compared to a uniform distribution. As the CDF of our clusters is below the uniform distribution, \textsc{MultiCAM}'s scatter is larger than its errors, meaning that its scatter is a conservative estimator of its errors.

To quantify this difference, we use the Kolmogorov--Smirnov (KS) test statistic, and find a difference between our $p$-value distribution and a uniform which could only occur 2\% of the time under the null hypothesis. Most of the difference appears to be constrained to very low $p$ values, meaning that MultiCAM is somewhat overestimating the error on its $\approx30{-}40\%$ most-accurate predictions.

\setcounter{figure}{0}
\section{Sensitivity of Merger Classification to Time Window Size}
\label{app:merger_widths}

\begin{figure}[h]
\centering
\includegraphics[width=\columnwidth]{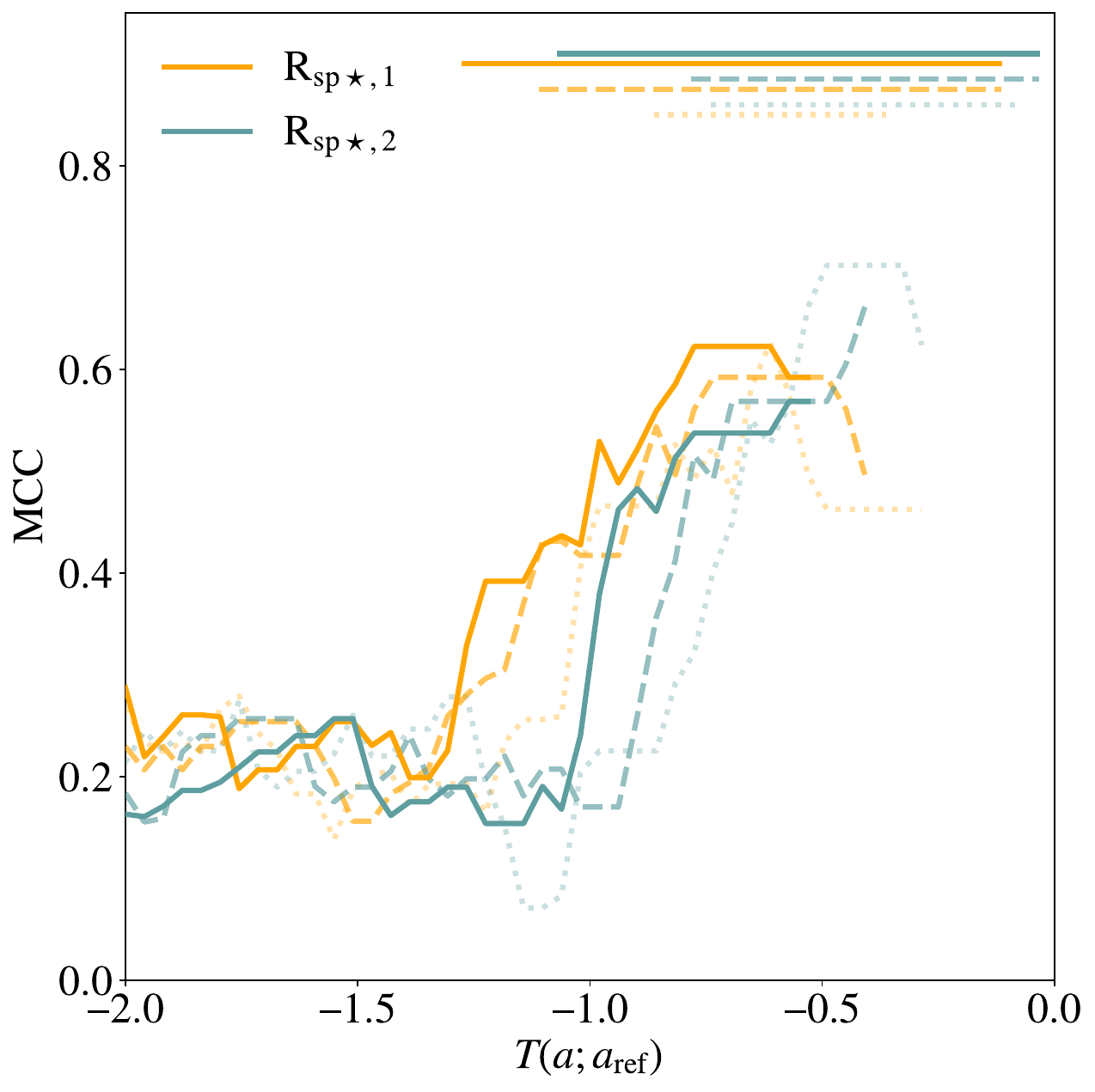}
    \caption{The optimized MCC classifying present-day $R_{\rm sp \star,1}$ and $R_{\rm sp \star,2}$ values in clusters that did or did not merge in time windows around $T$ of width 0.5, 0.75, 1.0 dynamical times. Curves corresponding to merger time window widths of 0.5, 0.75, and 1.0 are dotted, dashed, and solid lines, respectively. They also decrease in transparency. Horizontal line segments at the top indicate the time windows of mergers determined to be the best classified by the peaks of the MCC curves. For $R_{\rm sp \star,1}$, window widths of 0.75 and 1.0 result in particularly similar curves. For $R_{\rm sp \star,2}$, window widths of 0.5 and 0.75 result in similarly close curves that reach a similar peak.}
    \label{fig:merger_widths}
\end{figure}

In this section, we test how the present-day stellar splashback radii classify merger activity that occurred in windows of time of different widths. We test three window widths in units of the number of dynamical times: 0.5, 0.75, 1.0 with 1.0 being the fiducial width. In \autoref{fig:merger_widths}, we plot the optimized MCC classifying present-day $R_{\rm sp \star,1}$ and $R_{\rm sp \star,2}$ values in clusters that did or did not merge in time windows around $T$ of width 0.5, 0.75, 1.0 dynamical times (dotted, dashed, and solid lines, respectively). This figure is analogous to \autoref{fig:classification} and can be read in the same manner. The horizontal line segments at the top span the windows of time where the occurrence or non-occurrence of mergers is best classified. The centers of the line segments are therefore placed at the peaks of the MCC curves. Due to the flat peak in the dashed line of $R_{\rm sp \star,1}$ when classifying with windows of width 0.75 dynamical times, the line segment best probed indicates little difference between a window width of 1.0 and 0.75 dynamical times. A window width of 0.5 in the dotted line shows a similarly located peak but reaches slightly lower MCC values at most $T$. The flatness of the peak for $R_{\rm sp \star,2}$ and a width of 0.5 leads to very similar timescales probed by merger time windows of width 0.5 and 0.75. A window width of 1.0 for $R_{\rm sp \star,2}$ remains at lower MCC values as compared to 0.5 and 0.75, hinting that $R_{\rm sp \star,2}$ best probes shorter merger time windows in addition to more recent times. However, a detailed study of the window sizes best probed by the different cluster properties is out of the scope of this work and limited cluster sample. We choose to move forward with a merger time window of 1.0 dynamical time for all the cluster properties in our main analysis in \autoref{fig:classification}. We stress that the choice of width between 1.0 and 0.75 dynamical times does not impact the peak timescales or the peak MCC value found when classifying mergers in the multivariable space of $R_{\rm sp \star,1}$ and $R_{\rm sp \star,2}$ in \autoref{subsec:2D}.

\setcounter{figure}{0}
\section{Surface Brightness Profile Estimation}
\label{app:SB}
In this section, we describe our method for estimating the projected surface brightness profiles for each simulated cluster given its 3D stellar mass distributions as shown in \autoref{fig:SB}. 

We measure the 2D projected stellar mass density profiles for each cluster with the angular median method. This is done analogously to the 3D profiles described in \autoref{subsec:MeasureRsp} but from an arbitrary viewing angle, the $z$-direction. As discussed in \autoref{subsec:MeasureRsp}, we remove all infalling particles before creating profiles and measuring the splashback radii. This is necessary because our particle-tagging technique places sharp features at $R_{\rm vir}$ in the infalling particle population. We then add back in an estimate of the infalling population of star particles. We measure the 3D stellar mass density profile of infalling particles within $R_{\rm vir}$ and fit a power-law which will be extrapolated out to larger radii. Due to low number statistics of infalling particles in the inner regions of some of the clusters, we perform the power-law fit on a restricted radial range of the profile, chosen by eye for visibly affected clusters to avoid fitting to the affected regions. We emphasize that our primary goal is estimating the surface brightness of the projected ICL at the stellar splashback radii which appears negligibly affected by the details of the infall power law included. Assuming spherical symmetry, we integrate this 3D power-law infall profile over the z-direction, to obtain a 2D projected infall stellar mass density profile. We then combine this profile with the 2D projected profile of orbiting particles to obtain a total 2D projected stellar mass density profile.

In order to convert this stellar mass surface density profile into a surface brightness profile, we must choose a mass-to-light ratio that reasonably describes the population of stars in question. We choose to use mass-to-light versus color relations (MLCRs) to determine a reasonable range of mass-to-light ratios. Using clusters from the Dark Energy Survey (DES) and Atacama Cosmology Telescope (ACT) overlap, \citet{GoldenMarx2023} measured the $r-i$ color of the ICL over a redshift range $z = 0.2 - 0.8$. Using their Figure 10, we estimate that the median color in their outer radial bin ($\sim$200 kpc) for clusters at lookback times between 3.0 Gyr and 4.0 Gyr to be $r-i = 0.4$ with bootstrapped standard deviation on the median reaching $0.3$ and $0.5$. As we aim to compare to the surface brightness limits of the Vera C. Rubin Observatory's LSST, the Nancy Grace Roman Space Telescope, and Euclid, we estimate mass-to-light ratios in bands relevant to each survey.

For LSST, we estimate the $r$-band mass-to-light ratio using the MLCR linear fits provided by \citet{RoedigerCourteau2015}:
\begin{equation}
\log_{10}\left(\frac{M_\star/L_r}{M_\odot/L_\odot}\right) = -1.170 + 4.107\cdot(r-i).
\end{equation}
This fit was obtained by using the \textsc{Magphys} software package to apply the \citet{BC03} stellar population synthesis model to the joint prior model distribution described in \citet{Magphys2011}. For Roman and Euclid, we construct the MLCRs following the same procedure described in \citet{RoedigerCourteau2015} to derive fits for Roman's $H$ (F158) band and for Euclid's $H_{E}$ band after adding the appropriate transmission curves to \textsc{Magphys}. We obtain the following fits:

\vspace{1em}
\begin{align}
\log_{10}\left(\frac{M_\star/L_{H}}{M_\odot/L_\odot}\right) &= -1.222 + 2.188\cdot(r-i) \\
\log_{10}\left(\frac{M_\star/L_{H_E}}{M_\odot/L_\odot}\right) &= -1.274 + 2.125\cdot(r-i).
\end{align}

For each survey, we determine the mass-to-light ratios in the relevant band corresponding to our fiducial color, $r-i = 0.4$ and our lower and upper bounds, $r-i = 0.3$ and $r-i = 0.5$. These $M$/$L$ for each survey are presented in \autoref{tab:M_L}.
\begin{table}[ht]
\centering
\begin{tabular}{|l|l|l|l|}
\hline
\textbf{$r-i$ color} & \textbf{LSST $M$/$L_{r}$} & \textbf{Euclid $M$/$L_{H_E}$} & \textbf{Roman $M$/$L_{H}$} \\ \hline
0.3 & 1.11 & 0.231 & 0.272 \\ \hline
0.4 (fiducial) & 2.39 & 0.377 & 0.450 \\ \hline
0.5 & 5.13 & 0.615 & 0.746 \\ \hline
\end{tabular}
\caption{Table recording the $M$/$L$ values calculated in each survey band for each $r - i$ color using the obtained MLCR relations.}
\label{tab:M_L}
\end{table}

After using these mass-to-light ratios to obtain luminosity surface density profiles, we incorporate the effect of redshift dimming and convert to surface brightness profiles. The resulting medians of these surface brightness profiles for our fiducial, lower bound, and upper bound on the mass-to-light ratio are shown in \autoref{fig:SB} and discussed in \autoref{sec:obs}.

\bibliography{references}{}
\bibliographystyle{aasjournal}

\end{document}